\documentclass[aps,pre,preprint,superscriptaddress,amsmath,amsfonts,showpacs]{revtex4-1}
\usepackage{amsmath}
\usepackage{amssymb}
\usepackage{graphicx}
\usepackage{subfigure}
\usepackage{mathrsfs}
\usepackage{hyperref}

\begin{document}

\title{Homoclinic orbits and chaos in a pair of parametrically-driven
  coupled nonlinear resonators}%
\author{Eyal Kenig}
\affiliation{Raymond and Beverly Sackler School of Physics and
  Astronomy, Tel Aviv University, Tel Aviv 69978, Israel}
\author{Yuriy A.~Tsarin}
\affiliation{Institute of Radio Astronomy, National Academy of
  Sciences of Ukraine, 4 Krasnoznamennya St., Kharkov 61002,
  Ukraine} 
\author{Ron Lifshitz}
\email[Corresponding author:\ ]{ronlif@tau.ac.il}
\affiliation{Raymond and Beverly Sackler School of Physics and
  Astronomy, Tel Aviv University, Tel Aviv 69978, Israel}

\begin{abstract}
  We study the dynamics of a pair of parametrically-driven coupled
  nonlinear mechanical resonators of the kind that is typically
  encountered in applications involving microelectromechanical and
  nanoelectromechanical systems (MEMS \& NEMS). We take advantage of
  the weak damping that characterizes these systems to perform a
  multiple-scales analysis and obtain amplitude equations, describing
  the slow dynamics of the system. This picture allows us to expose
  the existence of homoclinic orbits in the dynamics of the integrable
  part of the slow equations of motion.  Using a version of the
  high-dimensional Melnikov approach, developed by Kova\v{c}i\v{c} and
  Wiggins [Physica D, \textbf{57}, 185 (1992)], we are able to obtain
  explicit parameter values for which these orbits persist in the full
  system, consisting of both Hamiltonian and non-Hamiltonian
  perturbations, to form so-called \v{S}ilnikov orbits, indicating a
  loss of integrability and the existence of chaos. Our analytical
  calculations of \v{S}ilnikov orbits are confirmed numerically.
\end{abstract}
\date{July 22, 2010}


\maketitle

% - -  - - - - - - - - - - - - - - - - - - - - - - - - - - - - - - - -

\section{INTRODUCTION}

Microelectromechanical system (MEMS) and nanoelectromechanical systems
(NEMS) have been attracting much attention in recent
years~\cite{RoukesPlenty,Cleland03,C00}.  MEMS \& NEMS resonators are
typically characterized by very high frequencies, extremely small
masses, and weak damping. As such, they are naturally being developed
for a variety of applications such as sensing with unprecedented
accuracy~\cite{Rugar04,Ilic04,Yang06,*Li07,*Naik09}, and also for
studying fundamental physics at small scales---exploring mesoscopic
phenomena~\cite{Schwab00,Weig04} and even approaching quantum
behavior~\cite{LaHaye,*Naik06,*Rocheleau10,OConnell10}.  MEMS \& NEMS
resonators often exhibit nonlinear behavior in their
dynamics~\cite{LCreview,Rhoads10}. This includes nonlinear resonant
response showing frequency pulling, multistability, and
hysteresis~\cite{C00,turner98,zaitsev,aldridge05,kozinsky07}, as well
as the formation of extended~\cite{BR} and
localized~\cite{sato06,*sato07,*sato08} collective states in arrays of
coupled nonlinear resonators, and the appearance of chaotic
dynamics~\cite{scheible02,Demartini07,Karabalin09}. Nonlinearities are
often a nuisance in actual applications, and schemes are being
developed to avoid them~\cite{Kacem09,*Kacem10}, but one can also
benefit from the existence of nonlinearity, for example in
mass-sensing applications~\cite{zhang02,Buks06}, in achieving
self-synchronization of large arrays~\cite{sync1,*sync2}, and even in
the observation of quantum behavior~\cite{katz07,*katz08}.

MEMS \& NEMS offer a wonderful experimental testing ground for
theories of chaotic dynamics. Numerical investigations of a number of
models of MEMS \& NEMS resonators have demonstrated period-doubling
transitions to chaos~\cite{Liu04,De05,Park08,Karabalin09,Haghighi10},
yet there are very few analytical results. One of the simplest models
of chaotic motion is that of the Duffing resonator with a double-well
potential, described by the Hamiltonian
\begin{equation}
  \label{eq:duffing}
  H(x,p) = \frac1{2m} p^2 + \frac12 k x^2 + \frac14 \alpha x^4,
\end{equation}
with $k<0$ and $\alpha>0$. This simple mechanical system has a
homoclinic orbit for $H=0$, connecting a saddle at the origin of phase
space to itself. Upon the addition of damping and an external drive
this system develops a particular kind of chaotic motion called
\emph{horseshoe chaos}~\cite{wiggins}, which can be studied
analytically using the Melnikov approach~\cite{Melnikov63}. The stable
manifold leading into the saddle and the unstable manifold leading
away from the saddle, which coincide in the unperturbed
Hamiltonian~(\ref{eq:duffing}), are deformed when damping and a drive
are added. Yet, conditions can be found analytically using the
Melnikov function, which measures the distance between the two
manifolds, under which they intersect transversely leading to the
possibility of observing chaotic dynamics. What one observes in
practice is a random-like switching of the resonator between the two
wells. Thus, having an analytical criterion for asserting the
existence of chaotic motion allows one to distinguish it from random
stochastic motion that might arise from noise.  Such a Hamiltonian as
in Eq.~(\ref{eq:duffing}) was implemented in a MEMS device using an
external electrostatic potential by DeMartini \emph{et
  al.}~\cite{Demartini07}, and studied using the Melnikov approach.

Here we wish to study the possibility of observing horseshoe chaos in
typical NEMS resonators, which are described by a potential as in
Eq.~(\ref{eq:duffing}), but of an elastic origin with $k$ and $\alpha$
both positive. Individual resonators of this type do not exhibit
homoclinic orbits for any value of $H$, and therefore are not expected
to display horseshoe chaos under a simple periodic drive.
Nevertheless, a pair of coupled resonators of this kind---like the
ones studied experimentally by Karabalin \emph{et
  al.}~\cite{Karabalin09}---are shown below to possess homoclinic
orbits in their collective dynamics, and are therefore amenable to
analysis based on a high-dimensional version of the Melnikov
approach~\cite{Holmes82a,*Holmes82b}. We employ here a particular
method, developed by Kova\v{c}i\v{c} and
Wiggins~\cite{kovacic1,*kovacic2,*kovacic3}, which is a combination of
the high-dimensional Melnikov approach and geometric singular
perturbation theory. This method enables us to find conditions, in
terms of the actual physical parameters of the resonators, for the
existence of an orbit in 4-dimensional phase space, which is
homoclinic to a fixed point of a saddle-focus type.  Such an orbit,
called a \v{S}ilnikov orbit~\cite{silnikov}, provides a mechanism for
producing chaotic dynamics~\cite{wiggins}.

We study here the case of parametric, rather than direct driving, but
this is not an essential requirement of our analysis. On the other
hand, having weak damping, or a large quality factor, characteristic
of typical MEMS \& NEMS resonators, is essential for the analysis that
follows. First of all, as was demonstrated in a number of earlier
examples~\cite{LC,BCL,*kenig1,*kenig2}, it leads to a clear separation
of time scales---a fast scale defined by the high oscillation
frequencies of the resonators, and a slow scale defined by the damping
rate. This allows us to perform a multiple-scales analysis in
Sec.~\ref{amp eq} and obtain amplitude equations to describe the slow
dynamics of the system of coupled resonators. It is in the slow
dynamics that the homoclinic orbits are found. Secondly, the weak
damping, which requires only a weak drive to obtain a response, allows
us to treat both the damping and the drive as perturbations, even with
respect to the slow dynamics. Therefore, in Sec.~\ref{unper sys} we
set the parametric drive amplitude and the damping to zero in the
amplitude equations, which makes them integrable. This allows us, in
Sec.~\ref{analytical homoclinic}, to find conditions for the existence
of homoclinic orbits and to obtain analytical expressions for these
orbits. We emphasize that these orbits reside in a 4-dimensional phase
space, and as such are homoclinic not to a point, but rather to a
whole invariant 2-dimensional manifold in the shape of a semi-infinite
cylinder. Of these, we identify a subset of orbits, satisfying a
particular resonance condition, that are precisely heteroclinic,
connecting pairs of points in 4-dimensional phase space. In
Sec.~\ref{intersections} we reintroduce the drive and the damping into
the equations as small perturbations, and use the high-dimensional
Melnikov method to determine which of the heteroclinic orbits,
determined through the resonance condition in the unperturbed system,
survives under the perturbation. In Sec.~\ref{resonance} we study the
effects of the perturbation on the dynamics within the invariant
semi-infinite cylinder near the resonance condition.  Finally, in
Sec.~\ref{silnikov} we put everything together by calculating the
parameter values for which the end-points of the unperturbed
heteroclinic orbits are deformed in the perturbed system in such a way
that they become connected through the dynamics on the semi-infinite
cylinder, producing \v{S}ilnikov orbits, homoclinic to a fixed point
of a saddle-focus type. We conclude by verifying our analytical
calculation using numerical simulations. Our analysis implies that
conditions exist in the coupled resonator system that could lead to
chaotic motion.

\section{NORMAL MODE AMPLITUDE EQUATIONS}
\label{amp eq}

We consider a pair of resonators modeled by the equations of motion
\begin{align}\label{eom}
\ddot{u}_{n}  + u_{n} &+ u_{n}^{3} - \frac{1}{2}Q^{ - 1}(\dot{u}_{n-1} -
2\dot{u}_{n} + \dot{u}_{n+1})\nonumber\\
 &+ \frac{1}{2}\left[D + H\cos\omega_p t\right](u_{n-1} -  2u_{n} +
 u_{n+1}) = 0, 
\qquad n=1,2,
\end{align}
where $u_{n}$ describes the deviation of the $n^{th}$ resonator from
its equilibrium, and we label two fictitious fixed resonators as
$u_{0} = u_{3} = 0$ for convenience.  Detailed arguments for the
choice of terms introduced into these equations of motion are
discussed by Lifshitz and Cross~\cite{LC}, who modeled the particular
experimental realization of Buks and Roukes~\cite{BR}, although other
variations are possible~\cite{LCreview}.  The terms include an elastic
restoring force as in Eq.~(\ref{eq:duffing}) with positive linear and
cubic contributions (whose coefficients are both scaled to 1), a dc
electrostatic nearest-neighbor coupling term with a small ac component
responsible for the parametric excitation (with coefficients $D$ and
$H$ respectively), and a linear dissipation term, which is taken to be
of a nearest neighbor form, motivated by the experimental
indication~\cite{BR} that most of the dissipation comes from the
electrostatic interaction between neighboring beams. Note that the
electrostatic attractive force acting between neighboring beams decays
with the distance between them, and thus acts to slightly soften the
otherwise positive elastic restoring force. Lifshitz and Cross also
considered an additional nonlinear damping term, which we neglect here
for the sake of simplicity.  The resonators' quality factor $Q$ is
typically high in MEMS \& NEMS devices, which can be used to define a
small expansion parameter $\epsilon\ll1$, by taking $Q^{ - 1} =
\epsilon{\hat{\gamma}}$, with $\hat\gamma$ of order unity.  The drive
amplitude is then expressed as $H = \epsilon\hat{h}$, in anticipation
of the fact that parametric oscillations at half the driving frequency
require a driving amplitude which is of the same order as the linear
damping rate~\cite{LCreview}.

Following Lifshitz and Cross~\cite[Appendix B]{LC}, we use multiple
time scales to express the displacements of the resonators as
\begin{equation}\label{normal modes}
    x_{1,2}(t)  = 
     \frac{\sqrt{3\epsilon}}{2} \left(A_{1}(T)e^{i\omega_{1}t}\pm
     A_{2}(T)e^{i\omega_{2}t} + c.c.\right)
     +  \epsilon^{3/2}x_{1,2}^{(1)}(t) + ...,
\end{equation}
where $x_{1}$ is taken with the positive sign and $x_{2}$ with the
negative sign; with a slow time $T = \epsilon t$, and where the normal
mode frequencies are given by $\omega_{1}^{2} = 1 - D/2$, and
$\omega_{2}^{2} = 1 - 3D/2$.  Substituting Eq.~(\ref{normal modes})
into the equations of motion~(\ref{eom}) generates secular terms that
yield two coupled equations for the complex amplitudes $A_{1,2}$. If we
measure the drive frequency relative to twice $\omega_{2}$ by setting
$\omega_p = 2\omega_2 + \epsilon\Omega$, express $\omega_{1}$ relative
to $\omega_{2}$ as $\omega_{1} = \omega_{2} + 2\epsilon\Omega_{1}$,
and express the complex amplitudes using real amplitudes and phases as
\begin{eqnarray}
A_{1}(T)& = &a_{1}(T)e^{i[\chi_{1}(T) + (\Omega/2 - 2\Omega_{1})T]},\nonumber\\
A_{2}(T)& = &a_{2}(T)e^{i[\chi_{2}(T) + \Omega T/2]},
\end{eqnarray}
the real and imaginary parts of the two secular amplitude equations
become
\begin{subequations}
\label{mode equations}
\begin{eqnarray}
  \frac{da_{1}}{dT} & = & -\frac{1}{4}\hat{\gamma}a_{1} -
  \frac{\hat{h}}{8\omega_{1}}a_{1}\sin2\chi_{1}
  - \frac{9}{8\omega_{1}}a_{2}^{2}a_{1}\sin2(\chi_{2} -
  \chi_{1}),\\
   \frac{d\chi_{1}}{dT} & = & 2\Omega_1 - \frac12\Omega -
  \frac{\hat{h}}{8\omega_{1}}\cos2\chi_{1}
    + \frac{9}{8\omega_{1}}\bigl[a_{1}^{2} + 2a_{2}^{2} +
  a_{2}^{2}\cos2(\chi_{2} - \chi_{1})\bigr],\\
  \frac{da_{2}}{dT} & = & -\frac{3}{4}\hat{\gamma} a_{2} -
  \frac{3\hat{h}}{8\omega_{2}}a_{2}\sin2\chi_{2}
   - \frac{9}{8\omega_{2}}a_{1}^{2}a_{2}\sin2(\chi_{1} -
  \chi_{2}),\\
   \frac{d\chi_{2}}{dT} & = &  - \frac12\Omega -
  \frac{3\hat{h}}{8\omega_{2}}\cos2\chi_{2}
    + \frac{9}{8\omega_{2}}\bigl[a_{2}^{2} + 2a_{1}^{2} +
  a_{1}^{2}\cos2(\chi_{1} - \chi_{2})\bigr].
\end{eqnarray}
\end{subequations}
Steady-state solutions, oscillating at half the parametric drive
frequency, are obtained by setting $da_{i}/dT = d\chi_{i}/dT = 0$ in
Eqs.~(\ref{mode equations}) and solving the resulting algebraic
equations. We are interested in extending the investigation of these
amplitude equations. In particular, we want to identify the conditions
under which they may display chaotic dynamics. We should note that
equations similar to~(\ref{mode equations}) were also used for
modeling a variety of parametrically driven two-degree of freedom
systems such as surface waves in nearly-square tanks or vibrations of
nearly-square thin plates or of beams with nearly-square cross
sections~\cite{meron861,*meron862,feng89,*feng931,*fandw,*feng95,%
  zhang,moehlis09}.

\section{UNPERTURBED EQUATIONS---SETTING DAMPING AND DRIVE TO ZERO}
\label{unper sys}

We first consider the integrable parts of Eqs.~(\ref{mode equations}),
obtained by setting $\hat{\gamma} = \hat{h} = 0$, which after a
rescaling of the amplitudes, $a_{1}\rightarrow
a_{1}\sqrt{\omega_{2}8/9}$, and $a_{2}\rightarrow
a_{2}\sqrt{\omega_{1}8/9}$, become
\begin{subequations}
\label{hamiltonian scaled}
\begin{eqnarray}
  \frac{da_{1}}{dT}& = &a_{2}^{2}a_{1}\sin2(\chi_{1} - \chi_{2}),\\
  \frac{d\chi_{1}}{dT} & = &  - \left(\frac{\Omega}{2} -
  2\Omega_{1}\right) + a_{2}^{2}\left[2 + \cos2(\chi_1 -
  \chi_2)\right] + \frac{\omega_{2}}{\omega_{1}}a_{1}^{2},\\
\frac{da_{2}}{dT}& = & - a_{1}^{2}a_{2}\sin2(\chi_{1} - \chi_{2}),\\
   \frac{d\chi_{2}}{dT}& = &  - \frac{\Omega}{2} + a_{1}^{2}\left[2 +
  \cos2(\chi_1 - \chi_2)\right] + \frac{\omega_{1}}{\omega_{2}}a_{2}^{2}.
\end{eqnarray}
\end{subequations}
In Sec.~\ref{intersections} we will reintroduce the driving and
damping terms as a perturbation. We transform Eqs.~(\ref{hamiltonian
  scaled}) into a more familiar form, which has been studied in the
context of higher dimensional Melnikov
methods~\cite{wiggins,kovacic1,kovacic3}, by changing to two pairs of
action-angle variables: (i) $B = a_{1}^2/2$, $\theta = \chi_{1} -
\chi_{2}$; and (ii) $I = (a_{1}^{2} + a_{2}^{2})/2$, $\phi =
\chi_{2}$. After defining $\delta = \omega_{1}/\omega_{2}$, and rescaling
time as $T\rightarrow T/2$, we obtain the unperturbed Hamilton
equations
\begin{subequations}
\label{radial}
\begin{eqnarray}
  \frac{dB}{dT} &=
  &-\frac{\partial\tilde{H}_{0}(B,\theta,I)}{\partial\theta} = 2B(I -
  B)\sin2\theta \label{hamilton5},\\ 
  \frac{d\theta}{dT} &= &\frac{\partial \tilde{H}_{0}(B,\theta,I)}{\partial
    B} =  \Omega_{1} + I(2 - \delta + \cos2\theta) -
  B\left(4 - \frac{\delta^{2} + 1}{\delta} +
    2\cos2\theta\right),\label{hamilton6}\\
  \frac{dI}{dT} & =
  &-\frac{\partial\tilde{H}_{0}(B,\theta,I)}{\partial\phi} =
  0,\label{hamilton7}\\ 
  \frac{d\phi}{dT} & =
  &\frac{\partial\tilde{H}_{0}(B,\theta,I)}{\partial I} = \delta I -
  \frac{\Omega}{4} + B(2 - \delta + \cos2\theta),\label{hamilton8} 
\end{eqnarray}
\end{subequations}
where the Hamiltonian $\tilde{H}_{0}$, which generates these
equations, is expressed as
\begin{equation}\label{hamiltonian2}
  \tilde{H}_{0}(B,\theta,I) = \frac{\delta I^{2}}{2} - \frac{\Omega}{4} I
  - B^{2}\left(2 - \frac{\delta^{2} + 1}{2\delta}\right)
  + B\left[I(2 - \delta) + \Omega_{1}\right] + B(I - B)\cos2\theta.
\end{equation}
Thus, both $I$ and $\tilde{H}_0$ are constants of the motion in the
unperturbed system. Note that $(B,\theta,I,\phi) \in \mathbb{R}^+
\times \mathbb{S} \times \mathbb{R}^+ \times \mathbb{S}$, where
$\mathbb{S}$ is the unit circle, and $\mathbb{R}^+$ are the
non-negative reals.

It is convenient to describe the dynamics also in terms of the
Cartesian variables $x = a_1 \cos(\chi_1-\chi_2) =
\sqrt{2B}\cos\theta$ and $y = a_1 \sin(\chi_1-\chi_2) =
\sqrt{2B}\sin\theta$, in place of $B$ and $\theta$, thereby obtaining
the Hamilton equations
\begin{subequations}
  \label{hamilton jacobi}
  \begin{eqnarray}
    \frac{dx}{dT} &= &-\frac{\partial H_{0}(x,y,I)}{\partial
      y} = y^{3}\left(1 - \frac{\delta^{2} +
        1}{2\delta}\right) + x^{2}y\left(2 - \frac{\delta^{2} +
        1}{2\delta}\right)
    - y\left[I(1 - \delta) + \Omega_{1}\right] \label{hamilton1},\\
    \frac{dy}{dT} &= &\frac{\partial
      H_{0}(x,y,I)}{\partial x} = - x^{3}\left(3 - \frac{\delta^{2} +
        1}{2\delta}\right) - y^{2}x\left(2 - \frac{\delta^{2} +
        1}{2\delta}\right)
    + x\left[I(3 - \delta) + \Omega_{1}\right] \label{hamilton2},\\
    \frac{dI}{dT} &= &-\frac{\partial H_{0}(x,y,I)}{\partial
      \phi} = 0\label{hamilton3},\\
    \frac{d\phi}{dT} &= &\frac{\partial
      H_{0}(x,y,I)}{\partial I} = \delta I - \frac{\Omega}{4} +
    \frac{x^{2}}{2}(3 - \delta) + 
    \frac{y^{2}}{2}(1 - \delta) \label{hamilton4},
  \end{eqnarray}
\end{subequations}
where $y$ plays the role of a coordinate and $x$ is its conjugate
momentum, and where the Hamiltonian $H_0$ is now given by
\begin{eqnarray}\label{hamiltonian}\nonumber
  H_{0}(x,y,I)& = &  \frac{\delta I^{2}}{2} - \frac{\Omega}{4} I -
  \frac{x^{4}}{4}\left(3 - \frac{\delta^{2} + 1}{2\delta}\right)
  - \frac{y^{4}}{4}\left(1 - \frac{\delta^{2} +
      1}{2\delta}\right) - x^{2}y^{2}\left(1 - \frac{\delta^{2} +
      1}{4\delta}\right)\\
  & + &\frac{x^{2}}{2}\left[I(3 - \delta) + \Omega_{1}\right] +
  \frac{y^{2}}{2}\left[I(1 - \delta) + \Omega_{1}\right].
\end{eqnarray}

\section{ANALYTICAL EXPRESSIONS FOR HOMOCLINIC ORBITS}
\label{analytical homoclinic}

We wish to identify the conditions under which there exist homoclinic
orbits in the unperturbed system. These orbits will potentially lead
to chaotic dynamics once we reintroduce the damping and the drive in
the form of small perturbations. We therefore consider the fixed point
$x = y = 0$ in the unperturbed $(x,y)$ plane, as given by
Eqs.~(\ref{hamilton1}) and (\ref{hamilton2}). A linear analysis of
this fixed point reveals that it is a saddle for values of the
positive constant of motion $I$, that satisfy the inequality
$[I(1-\delta)+\Omega_1][I(3-\delta)+\Omega_1]<0$. This implies that
the fixed point at $x=y=0$ is never a saddle if the fixed parameter
$\delta<1$; it is a saddle for $1<\delta<3$, if $I >
\Omega_1/(\delta-1)$; and it is a saddle for $\delta>3$, if
$\Omega_1/(\delta-1) < I < \Omega_1/(\delta-3)$.  We shall restrict
ourselves here to values $1<\delta<3$, therefore to obtain a saddle
one must only ensure that $I>\Omega_1/(\delta-1)$. In the full
four-dimensional system given by Eqs.~(\ref{hamilton jacobi}) this
saddle point describes a two-dimensional invariant semi-infinite 
cylinder, or annulus, 
\begin{equation}\label{invariant_manifold}
    \mathscr{M} = \left\{(x,y,I,\phi)\ \big|\ x = 0,\ y = 0,\
    \frac{\Omega_{1}}{\delta - 1} < I \right\},\qquad 1<\delta<3,
\end{equation}
where $\phi$ is unrestricted within the unit circle.
%% If $\delta > 3$ then $I_{2}$ is bounded from above by $\Omega_{1}/(\delta -
%% 3)$, but we restrict ourselves to the parameter values $1 < \delta <
%% 3$.  The reason for defining $\mathscr{M}$ with the boundaries $I_{1}$
%% and $I_{2}$ will be clarified later.  
The trajectories on $\mathscr{M}$ are periodic orbits given by $I =
\textmd{constant}$ and $\phi = (\delta I - \Omega/4)T + \phi_{0}$. For
the resonant value of $I\equiv I^{r} = \Omega/4\delta$ the rotation
frequency vanishes, and the periodic orbit becomes a circle of fixed
points. Of course, this trivial unperturbed dynamics on $\mathscr{M}$
undergoes a dramatic change under the addition of perturbations.

The two-dimensional invariant annulus $\mathscr{M}$ has
three-dimensional stable and unstable manifolds, denoted as
$W^{s}(\mathscr{M})$ and $W^{u}(\mathscr{M})$, respectively, which
coincide to form a three-dimensional homoclinic manifold $\Gamma\equiv
W^{s}(\mathscr{M})\cap W^{u}(\mathscr{M})$. Trajectories on the
homoclinic manifold $\Gamma$ are homoclinic orbits that connect the
origin of the $(x,y)$ plane to itself. Thus, the constant value of the
Hamiltonian along such an orbit is equal to its value at the origin,
namely $H_{0}(0,0,I) = H_{0}(x,y,I) = \tilde{H}_{0}(B,\theta,I)$,
which immediately yields an equation for the homoclinic orbits in
terms of the action-angle variables
\begin{equation}\label{B as theta}
    B^h(\theta,I) = \frac{2\delta\left[I\left(\delta - 2
    -\cos2\theta\right)  - \Omega_{1}\right]}{\delta^{2} -
    2\delta\left(2+\cos2\theta\right) + 1}.
\end{equation}
To obtain the temporal dependence of the dynamical variables along the
homoclinic orbit, we substitute the homoclinic orbit equation~(\ref{B
  as theta}) into Eq.~(\ref{hamilton6}), to get
\begin{equation}\label{theta dot}
    \frac{d\theta}{dT} = I(\delta-2 - \cos2\theta) -\Omega_1.
\end{equation}
Next, we note that $\chi_{1} = \phi + \theta$, and use the Hamiltonian
(\ref{hamiltonian2}) to get
\begin{equation}\label{psi dot}
    \frac{d\chi_{1}}{dT} = I\delta - \frac{\Omega}{4} + B\frac{1 -
    \delta^{2}}{2\delta}.
\end{equation}
We then integrate Eq.~(\ref{theta dot}), substitute the result into
Eq.~(\ref{B as theta}), and the latter into Eq.~(\ref{psi dot}), and
finally integrate Eq.~(\ref{psi dot}) to obtain analytical expressions
for the temporal dependence of the dynamical variables along orbits
that are homoclinic to $\mathscr{M}$.  

For $I >2\delta\Omega_{1}/(\delta^{2} - 1)$ we define $q\equiv
I(\delta^2 - 1) - 2\delta\Omega_1 > 0$, and find that
$\theta_0\equiv\theta(T=0)=0,\pi$, and that the homoclinic orbits are
given by
\begin{subequations}
\label{orbit}
\begin{eqnarray}
  B^{h}(T,I) &= &\frac{2\delta a^2}{q\cosh(2aT) + p},\label{B}\\
  \tan\left(\theta^{h}(T,I)\right) &= &-\sqrt{\frac{I(\delta - 3) -
      \Omega_{1}}{I(1 - \delta) + \Omega_{1}}}\tanh (aT),\label{theta}\\ 
  \chi_{1}^{h}(T,I) &= &-\frac{a(\delta^{2} - 1)}{\sqrt{p^{2} - q^{2}}} 
  \text{ arctanh}\left(\sqrt\frac{p-q}{p+q}\tanh aT\right) + (\delta I
  - \frac{\Omega}{4})T + \chi_1(0)\label{psi},\\  
  \phi^{h}(T,I) &= &\chi_{1}^{h}(T,I) - \theta^{h}(T,I),
\end{eqnarray}
\end{subequations}
where
\begin{eqnarray}\label{params}
  p &= &\Omega_{1}(\delta^2 - 4\delta + 1) - I(\delta^3 - 6\delta^{2} +
  7\delta -2),\nonumber\\
%q &= & -I - 2\Omega_{1}\delta + I\delta^{2},\nonumber\\
%c &= & - 2\delta(k^{4} - 2I\Omega_{1}(-2 + \delta) + I^{2}(3 -
%4\delta + \delta^{2})),\nonumber\\ 
  a^2 &= &-\Omega_1^{2} + 2I\Omega_{1}(\delta-2) - I^{2}(\delta-3)(\delta-1).  
\end{eqnarray}

For $I < 2\delta\Omega_{1}/(\delta^{2} - 1)$ we redefine $q\equiv
2\delta\Omega_1 - I(\delta^2 - 1) > 0$, and  find that
$\theta_0=\pm\pi/2$, and that the homoclinic orbits are given by
Eqs.~(\ref{orbit}), with Eq.~(\ref{theta}) replaced by
\begin{equation}\label{new theta}
  \cot\left(\theta^{h}(T,I)\right) = -\sqrt{\frac{I(1 - \delta)
      + \Omega_{1}}{I(\delta - 3) - \Omega_{1}}}\tanh (aT).
\end{equation}
Thus, exactly at $I = 2\delta\Omega_{1}/(\delta^{2} - 1)$ (or $q=0$)
there is a global bifurcation in which the homoclinic orbit rotates
through an angle of $\pi/2$.

\begin{figure}
\begin{center}
  \subfigure[$\ I=6$]{
    \includegraphics[width = 0.3\textwidth]{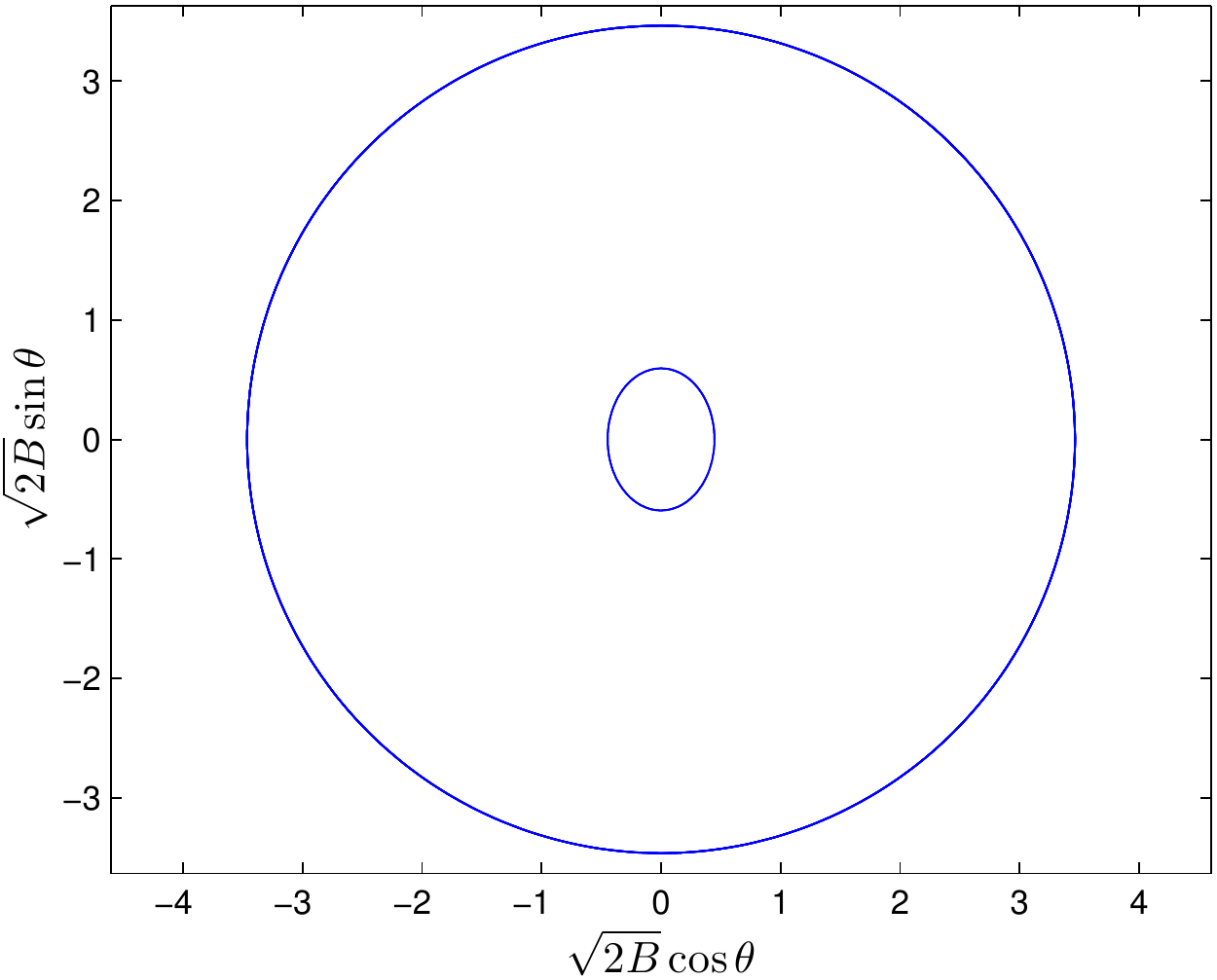}}
    \subfigure[$\ I=10$]{
   \includegraphics[width = 0.3\textwidth]{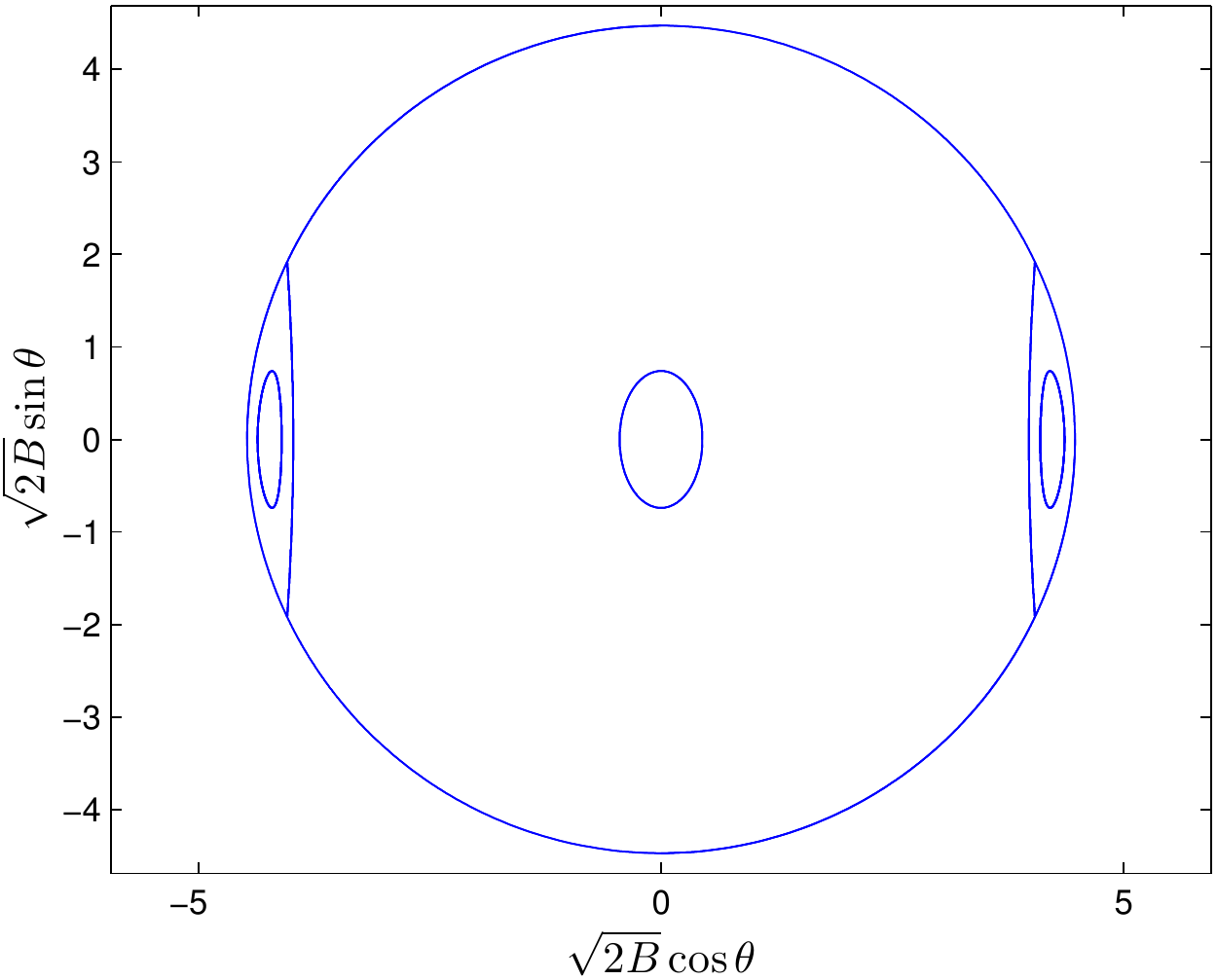}}
     \subfigure[$\ I=25$]{
  \includegraphics[width = 0.3\textwidth]{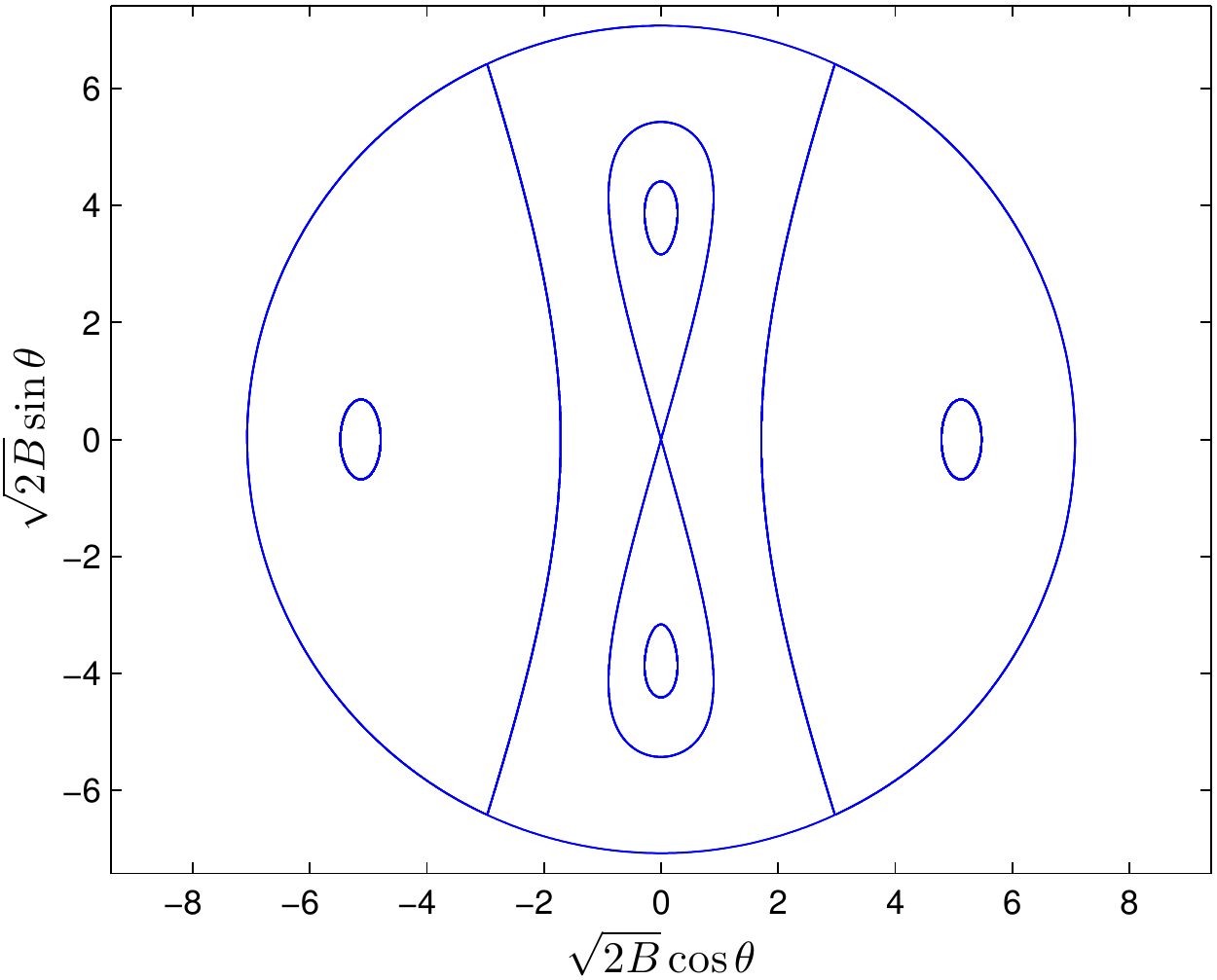}}
    \subfigure[$\ I=28.4267$]{
   \includegraphics[width = 0.3\textwidth]{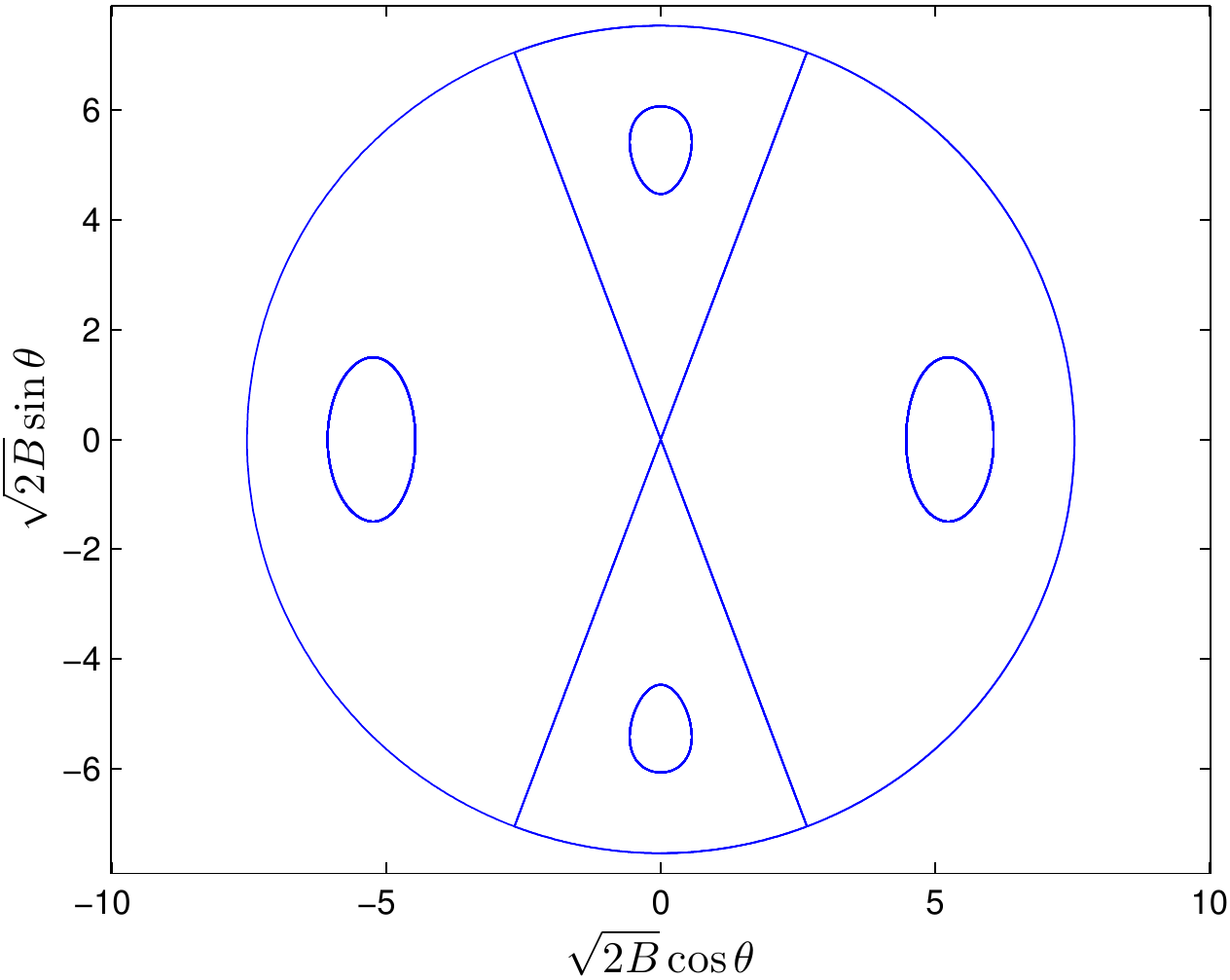}}
     \subfigure[$\ I=38.5$]{
  \includegraphics[width = 0.3\textwidth]{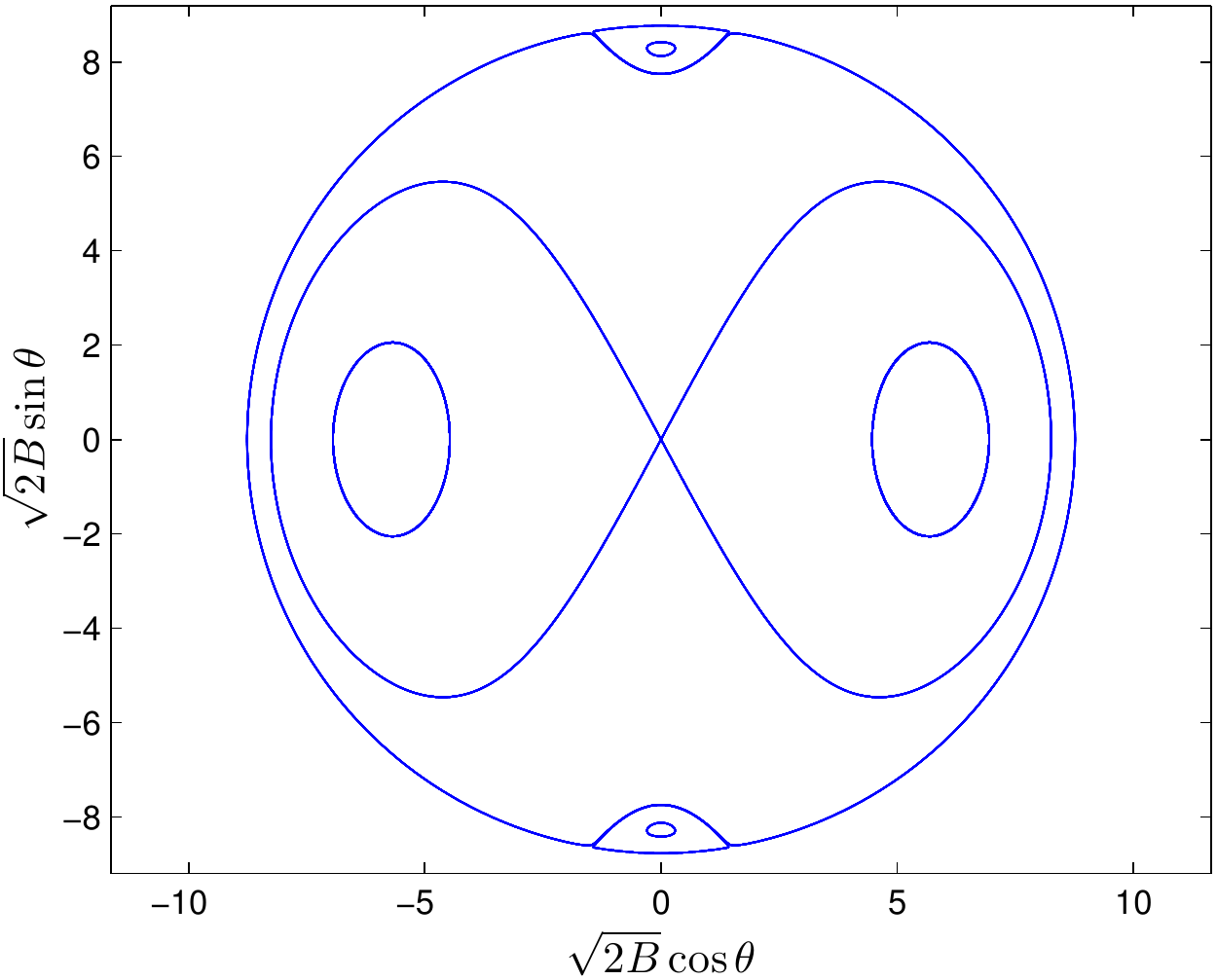}}
    \subfigure[$\ I=43$]{
   \includegraphics[width = 0.3\textwidth]{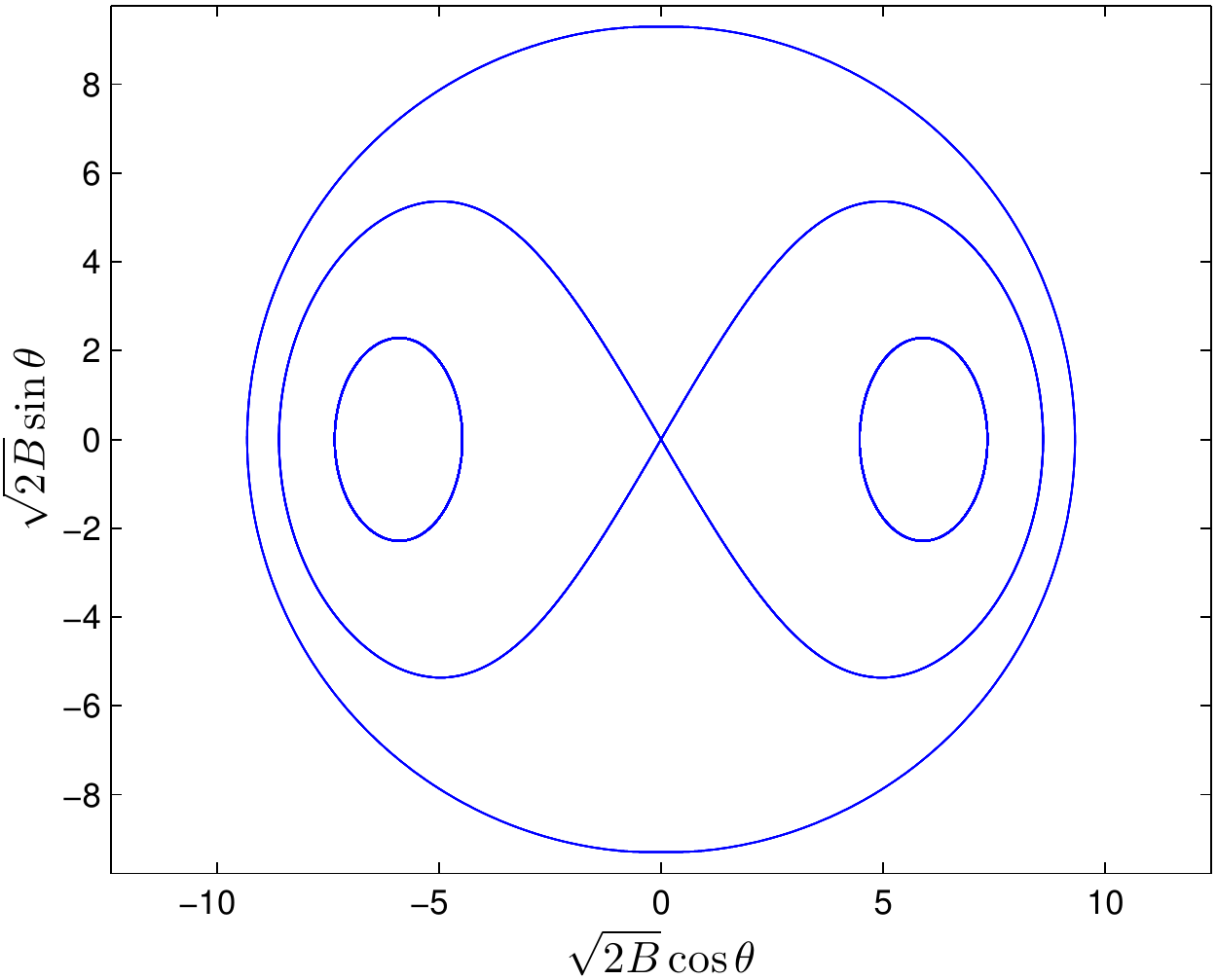}}
   \end{center}
   \caption{\label{phase_space} Numerical phase portraits of the unperturbed
     $(x,y)$ plane for $\delta = 2$, $\Omega_{1} = 21.32$, and
     different values of $I$ as noted in the individual captions. All
     figures show the trajectories for which $B$ has a fixed value
     equal to $I$. As explained in the text and for $\delta=2$, four
     hyperbolic fixed points appear on the $B=I$ circle for
     $2\Omega_1/5=8.528 < I < 2\Omega_1=42.64$; the $B=0$ fixed point
     changes from a center to a saddle at $I=\Omega_1=21.32$; and a
     global bifurcation rotating the homoclinic orbit through an angle
     of $\pi/2$, shown in panel (d), occurs at $I=4\Omega_1/3\simeq 28.4267$.}
\end{figure}

Some of the phase-space portraits of the unperturbed $(x,y)$ plane are
calculated numerically from Eqs.~(\ref{hamilton5}) and
(\ref{hamilton6}), for different values of $I$, and shown in
Fig.~\ref{phase_space}. From Eq.~(\ref{hamilton5}) it follows that the
value of $B$ is fixed if (a) $B=0$; or (b) $B=I$; or (c) $\theta$ is
an integer multiple of $\pi/2$ and $\theta$ is fixed.
Figures~\ref{phase_space}(a)--(f) all show the trajectories for which
$B$ has a fixed value equal to $I$. Four hyperbolic fixed points
appear on the $B=I$ circle for $\Omega_{1}/(3 - 1/\delta) \leq I \leq
\Omega_{1}(1 - 1/\delta)$, where solutions exist to the equation
$\partial\theta/\partial T = 0$ with $B$ replaced by $I$
[Figs.~\ref{phase_space}(b)--(e)]. As expected, the origin $B=0$ is
always a fixed point---a center for small values of $I$
[Figs.~\ref{phase_space}(a),(b)], which undergoes a pitchfork
bifurcation into a saddle when $\partial\theta/\partial T = 0$ with
$B=0$, occurring at $I= \Omega_{1}/(\delta - 1)$
[Figs.~\ref{phase_space}(c)--(f)]. Additional centers appear whenever
$\theta$ is an integer multiple of $\pi/2$ and solutions exist to the
equation $\partial\theta/\partial T = 0$ with $\cos2\theta=\pm1$
[Figs.~\ref{phase_space}(b)--(f)]. The global bifurcation at $I =
2\Omega_{1}\delta/(\delta^{2} - 1)$ where the homoclinic orbit rotates
by $\pi/2$ is shown in Fig.~\ref{phase_space}(d).

\begin{figure}
   \includegraphics[width = 0.7\textwidth]{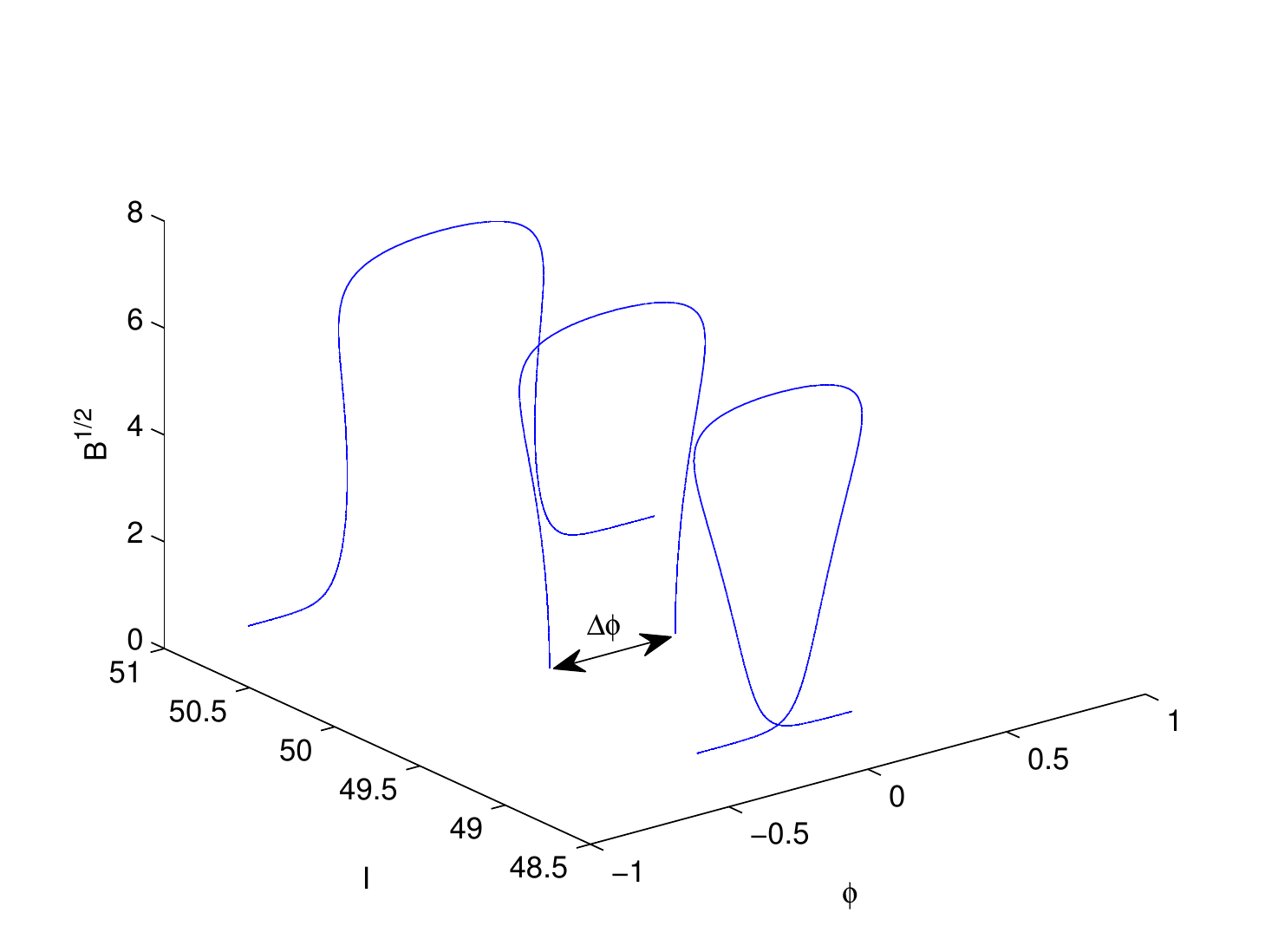}
   \caption{\label{unperturbed} Orbits homoclinic to
     $\mathscr{M}$. For $I = I^{r}$ (the orbit in the middle),
     $d\phi/dT = 0$ on $\mathscr{M}$, and the orbit is heteroclinic,
     connecting fixed points on $\mathscr{M}$ that are $\Delta\phi$
     apart. For $I\lessgtr I^{r}$, $d\phi/dT\lessgtr0$ on
     $\mathscr{M}$. The parameters are $\delta = 2,\Omega = 400,\Omega_{1}
     = 21.32$.}
\end{figure}

Note that we refer to the orbits given by Eqs.~(\ref{orbit}) as
homoclinic since they are homoclinic to $\mathscr{M}$. A few of these
orbits are shown in Fig.~\ref{unperturbed}. At resonance, for $I =
I^{r}$, the orbits are truly heteroclinic, connecting fixed points
that are $\Delta\phi$ apart, where $\Delta\phi = \Delta\chi_{1} -
\Delta\theta$, and
\begin{subequations}
\label{phase difference}
\begin{eqnarray}
  \Delta\theta & = & - 2\textmd{arctan}\sqrt{\frac{I^{r}(\delta
  - 3) - \Omega_{1}}{I^{r}(1 - \delta) + \Omega_{1}}},\\
  \Delta\chi_{1} & = &   - \frac{2a(\delta^{2} - 1)}{\sqrt{p^{2} -
  q^{2}}}\textmd{arctanh}\sqrt\frac{p-q}{p+q}, 
\end{eqnarray}
\end{subequations}
and where for any variable $f$, $\Delta f \equiv f(T=\infty) - f(T=
-\infty)$. Such an unperturbed heteroclinic orbit is shown in the
middle of Fig.~\ref{unperturbed}.

\begin{figure}
  \includegraphics[width = 0.7\textwidth]{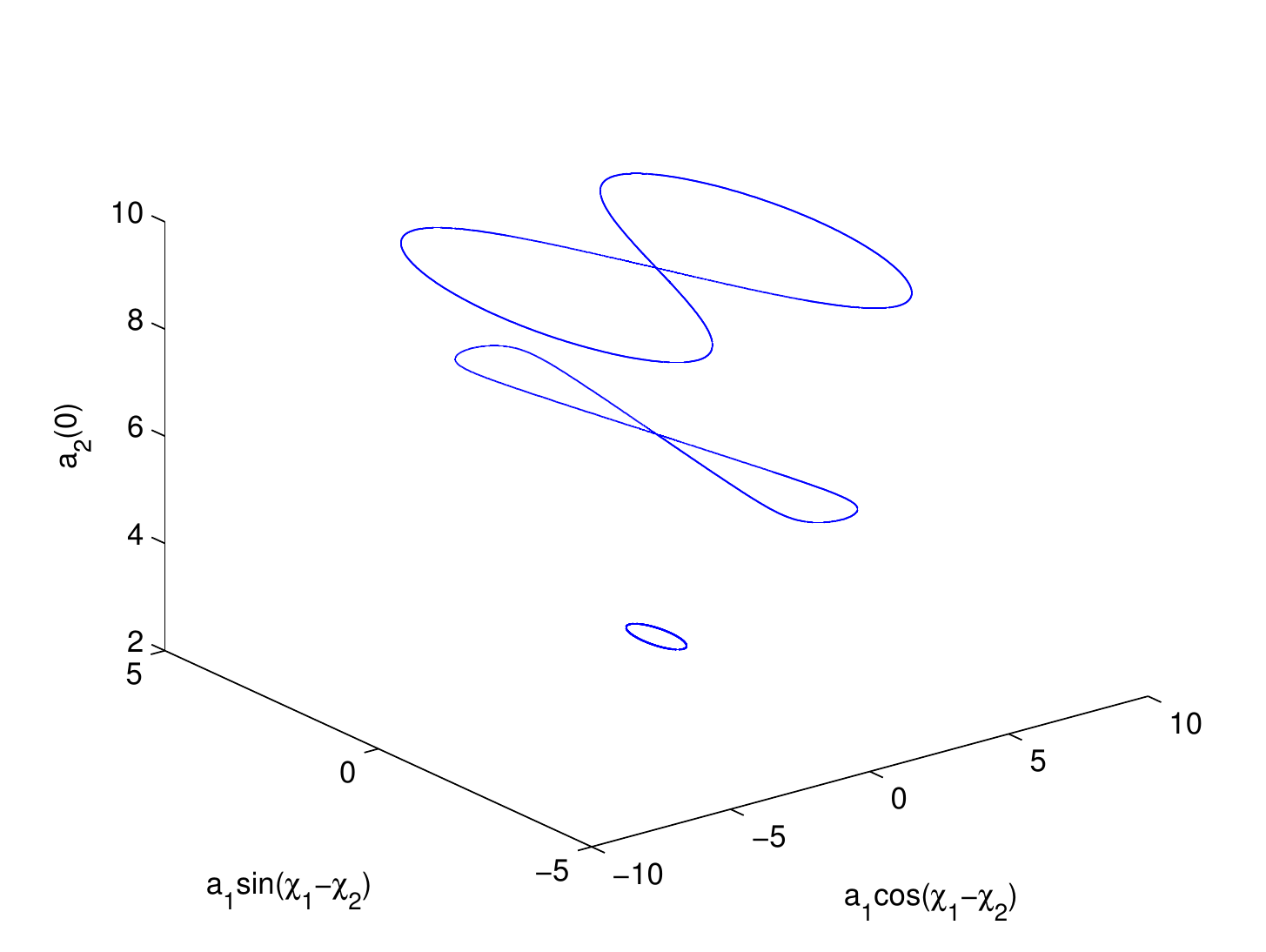}
  \caption{\label{homoclinic_simulation} Results of a numerical
    integration of Eqs.~(\ref{mode equations}) for different values of
    the initial amplitude of the second mode $a_{2}(0)$, with
    $\hat{h}=\hat{\gamma}=0$, $\omega_{1} = 0.8528$, $\omega_{2} =
    \omega_{1}/2$, $\Omega_{1} = 21.32$, and $\Omega = 400$. As
    expected, for $a_{2}(0) >
    \sqrt{16\Omega_{1}\omega_{1}\omega_{2}/9(\omega_{1} - \omega_{2})}
    \simeq 5.68$ the origin becomes a saddle, which rotates through an
    angle of $\pi/2$ at $a_{2}(0) =
    \sqrt{32\Omega_{1}\omega_{1}^{2}\omega_{2}/9(\omega_{1}^{2} -
      \omega_{2}^{2})} \simeq 6.46$.}
\end{figure}

We wish to demonstrate the results obtained so far also in terms of
original amplitude equations~(\ref{mode equations}) with
$\hat{h}=\hat{\gamma}=0$. The point $x = y = 0$ corresponds to $a_{1}
= 0$ in Eqs.~(\ref{mode equations}).  To start the simulation near
this point, we initiate the numerical solution with $a_{1}(0)\ll1$,
which through the definition of $I$ implies that
$a_{2}(0)\simeq\sqrt{16I\omega_{1}/9}$. The condition for having a
saddle at the origin of Eqs.~(\ref{hamilton1}) and (\ref{hamilton2}),
$I > \Omega_{1}/(\delta - 1)$, translates into the condition $a_{2}(0)
> \sqrt{16\Omega_{1}\omega_{1}/9(\delta - 1)} =
\sqrt{16\Omega_{1}\omega_{1}\omega_{2}/9(\omega_{1} - \omega_{2})}$.
The condition for the global bifurcation, rotating the homoclinic
orbit through $\pi/2$, given by $I=2\delta\Omega_1/(\delta^2-1)$,
translates into $a_{2}(0) =
\sqrt{32\Omega_{1}\omega_{1}^{2}\omega_{2}/9(\omega_{1}^{2} -
  \omega_{2}^{2})}$. These conditions are verified by a numerical
integration of Eqs.~(\ref{mode equations}) by varying the initial
amplitude of the out-of-phase mode, $a_{2}(0)$, as shown in
Fig.~\ref{homoclinic_simulation}.

\section{HOMOCLINIC INTERSECTIONS IN THE PERTURBED SYSTEM}
\label{intersections}

After having calculated the homoclinic orbits in the unperturbed
system, we now reintroduce the drive and the damping as perturbations
and study how they affect the dynamics. In particular, we want to
study the nature of the invariant annulus $\mathscr{M}$, and its
stable and unstable manifolds, $W^{s}(\mathscr{M})$ and
$W^{u}(\mathscr{M})$, under the perturbation, and use the Melnikov
criterion to find the conditions under which they can still intersect.
The perturbed equations are written in terms of the action-angle
variables as
\begin{subequations}
\label{per}
\begin{eqnarray}
 \frac{dB}{dT} &= &2B(I - B)\sin2\theta - \xi hB\sin2(\phi +
 \theta) - \xi\gamma B,\\
 \frac{d\theta}{dT} &= &\Omega_{1} + I(2 - \delta + \cos2\theta) -
 B\left(4 - \frac{\delta^{2} + 1}{\delta} +  2\cos2\theta\right)\nonumber\\
 &- &\frac{\xi h}{2}\left[\cos2(\phi + \theta) - 3\delta\cos2\phi\right],\\
 \frac{dI}{dT} &= &-\xi h\left[3\delta(I - B)\sin2\phi +
 B\sin2(\phi + \theta)\right] -\xi\gamma(3I - 2B)\label{per3},\\
\frac{d\phi}{dT} &= &- \frac{\Omega}{4} + \delta I + B(2 - \delta +
 \cos2\theta) - \frac{3\xi h\delta}{2}\cos(2\phi)\label{per4},
\end{eqnarray}
\end{subequations}
where we have quantified the perturbations by expressing the drive
amplitude and the damping as $\hat{h} = \xi8\omega_{1}h$ and
$\hat{\gamma} = \xi4\gamma$, respectively, where $\xi\ll1$ is a small
parameter.  It is instructive to write the perturbed system in the
general form
\begin{subequations}
\begin{eqnarray}
  \frac{dB}{dT} &= &-\frac{\partial
    \tilde{H}_{0}(B,\theta,I)}{\partial\theta} + \xi g^{B} 
  = -\frac{\partial \tilde{H}_{0}(B,\theta,I)}{\partial\theta} +
  \xi\left(-\frac{\partial
      \tilde{H}_{1}(B,\theta,I,\phi)}{\partial\theta} +
    d^{B}\right),\\
  \frac{d\theta}{dT} &= &\frac{\partial
    \tilde{H}_{0}(B,\theta,I)}{\partial B} + \xi g ^{\theta}
  = \frac{\partial \tilde{H}_{0}(B,\theta,I)}{\partial B} +
  \xi\frac{\partial \tilde{H}_{1}(B,\theta,I,\phi)}{\partial B},\\ 
  \frac{dI}{dT} & = & \xi g^{I} = \xi\left(-\frac{\partial
      \tilde{H}_{1}(B,\theta,I,\phi)}{\partial\phi} +
    d^{I}\label{dIdt}\right),\\ 
  \frac{d\phi}{dT} & = &   \frac{\partial \tilde{H}_{0}(B,\theta,I)}{\partial
    I} + \xi g^{\phi}\label{dphidt}
  = \frac{\partial \tilde{H}_{0}(B,\theta,I)}{\partial I} +
    \xi\frac{\partial \tilde{H}_{1}(B,\theta,I,\phi)}{\partial I}, 
\end{eqnarray}
\end{subequations}
where the perturbations due to the parametric drive are generated
from the Hamiltonian
\begin{equation}\label{H1Btheta}
   \tilde{H}_{1}(B,\theta,I,\phi) = -\frac{h}{2}\left[B\cos2(\phi +
   \theta) + 3\delta(I - B)\cos2\phi\right],
\end{equation}
and the dissipative perturbations are given by $d^{B} = - \gamma B$
and $d^{I} = - \gamma(3I - 2B)$.  Similarly, in terms of the Cartesian
variables, the perturbed system is written in this general form as
\begin{subequations}
\label{per_cartezian}
\begin{eqnarray}
  \frac{dx}{dT} &= &-\frac{\partial H_{0}(x,y,I)}{\partial y} + \xi
  g^{x}\label{dxdt}
  = -\frac{\partial H_{0}(x,y,I)}{\partial y} +
  \xi\left(-\frac{\partial H_{1}(x,y,I,\phi)}{\partial y} +
    d^{x}\right),\\
  \frac{dy}{dT} &= &\frac{\partial H_{0}(x,y,I)}{\partial x} + \xi
  g^{y}\label{dydt}
  = \frac{\partial H_{0}(x,y,I)}{\partial x} + \xi\left(\frac{\partial
  H_{1}(x,y,I,\phi)}{\partial x} + d^{y}\right),\\  
  \frac{dI}{dT} &= &\xi g^{I} = \xi\left(-\frac{\partial
      H_{1}(x,y,I,\phi)}{\partial\phi} + d^{I}\right),\\
  \frac{d\phi}{dT} &= &\frac{\partial H_{0}(x,y,I)}{\partial
    I} + \xi g^{\phi} = \frac{\partial H_{0}(x,y,I)}{\partial I} +
  \xi\frac{\partial  H_{1}(x,y,I,\phi)}{\partial I}, 
\end{eqnarray}
\end{subequations}
with
\begin{equation}
   H_{1}(x,y,I,\phi)= \frac{h}{4}\left\{\left[\left(3\delta-1\right)x^{2} +
   \left(3\delta+1\right)y^{2} - 6\delta I\right] \cos2\phi +
   2xy\sin2\phi\right\}, 
\end{equation}
and where $d^{x} = -\gamma x/2$, $d^{y} = -\gamma y/2$, and $d^{I} =
-\gamma(3I - x^{2} - y^{2})$. 

For $0 < \xi\ll1$ the unperturbed invariant annulus $\mathscr{M}$, and
its stable and unstable manifolds, $W^{s}(\mathscr{M})$ and
$W^{u}(\mathscr{M})$, persist as a \textit{locally invariant} annulus
$\mathscr{M}_{\xi}$ with stable and unstable manifolds,
$W^{s}(\mathscr{M}_{\xi})$ and $W^{u}(\mathscr{M}_{\xi})$
\cite{wiggins,kovacic1,kovacic2,kovacic3,kovacic96,feng931}. Due to the fact
that we use parametric rather than direct excitation, the point $x = y
= 0$ remains a fixed point of the perturbed Eq.~(\ref{dxdt}) and
(\ref{dydt}), so $\mathscr{M}_{\xi}$ is defined just like
$\mathscr{M}$ in Eq.~(\ref{invariant_manifold}).  However, the term
\textit{locally invariant} means that trajectories with initial
conditions on $\mathscr{M}_{\xi}$ may leave it through its lower
boundary at $I = \Omega_1/(\delta-1)$.  We want to find intersections
of the manifolds $W^{s}(\mathscr{M}_{\xi})$ and
$W^{u}(\mathscr{M}_{\xi})$, because such intersections may contain
orbits that are homoclinic to $\mathscr{M}_{\xi}$. This is done by
calculating the Melnikov integral, $M(I,\phi_{0})$, which is a measure
of the distance between these manifolds. If the Melnikov integral has
simple zeros [$M(I,\phi_{0}) = 0$ and $\partial
M(I,\phi_{0})/\partial\phi_{0} \neq 0$], the three-dimensional
manifolds $W^{s}(\mathscr{M}_{\xi})$ and $W^{u}(\mathscr{M}_{\xi})$
intersect transversely along two-dimensional surfaces.

The Melnikov integral is given by~\cite{wiggins,kovacic1,kovacic2,kovacic3}
\begin{equation}\label{Melnikov}
    M(I,\phi_{0}) = \int_{-\infty}^{\infty} \langle\textbf{n}(x^{h},y^{h},I),
    \textbf{g}(x^{h},y^{h},I,\phi^{h} + \phi_{0})\rangle dT, 
\end{equation}
where
\begin{eqnarray}\label{melnikov_integral}
  &&\textbf{n}(x,y,I) = \bigg(\frac{\partial H_{0}(x,y,I)}{\partial
  x},\frac{\partial H_{0}(x,y,I)}{\partial y}, 
  \frac{\partial H_{0}(x,y,I)}{\partial I} -\frac{\partial
  H_{0}(0,0,I)}{\partial I}\bigg),\\  
  &&\textbf{g}(x,y,I,\phi) = \left(g^{x},g^{y},g^{I}\right),
\end{eqnarray}
$x^{h}(T,I)$, $y^{h}(T,I)$, and $\phi^{h}(T,I)$ are the homoclinic
orbits given by Eqs.~(\ref{orbit}), and angular brackets denote the
standard inner product.  At resonance, the Melnikov integral
$M(I^{r},\phi_{0})$ can be calculated explicitly, because then
$\partial H_{0}(0,0,I^{r})/\partial I = 0$ and the integrand of the
Melnikov integral is given by
\begin{eqnarray}
  \langle\textbf{n},\textbf{g}\rangle &= &\frac{\partial
   H_{0}}{\partial x}g^{x} + \frac{\partial H_{0}}{\partial y}g^{y} +
   \frac{\partial H_{0}}{\partial I}g^{I}\nonumber\\ 
   &= &-\frac{\partial H_{0}}{\partial x}\frac{\partial
   H_{1}}{\partial y} + \frac{\partial H_{0}}{\partial
   y}\frac{\partial H_{1}}{\partial x} - \frac{\partial
   H_{0}}{\partial I}\frac{\partial H_{1}}{\partial \phi}
   + \frac{\partial H_{0}}{\partial x}d^{x} + \frac{\partial
   H_{0}}{\partial y}d^{y} + \frac{\partial H_{0}}{\partial I}d^{I}. 
\end{eqnarray}

For the unperturbed orbits we can use the chain rule and the fact that
$dI/dT = 0$ to obtain the relation
\begin{equation}\label{}
  \frac{dH_{1}}{dT}  = \frac{\partial H_{0}}{\partial x}\frac{\partial
    H_{1}}{\partial y} - \frac{\partial H_{0}}{\partial y}\frac{\partial
    H_{1}}{\partial x} + \frac{\partial H_{0}}{\partial
    I}\frac{\partial H_{1}}{\partial\phi},  
\end{equation}
so the Melnikov integrand reduces to
\begin{equation}\label{integrand}
  \langle\textbf{n},\textbf{g}\rangle =  - \frac{dH_{1}}{dT} +
  \frac{\partial H_{0}}{\partial x}d^{x} + \frac{\partial
  H_{0}}{\partial y}d^{y} + \frac{\partial H_{0}}{\partial I}d^{I}. 
\end{equation}
Upon transforming to the action-angle variables one has
\begin{equation}
  \frac{\partial H_{0}}{\partial x}d^{x} + \frac{\partial
  H_{0}}{\partial y}d^{y} = \frac{\partial \tilde{H}_{0}}{\partial
  B}d^{B},  
\end{equation}
and so the integrand~(\ref{integrand}) becomes 
\begin{eqnarray}\label{integrand2}
    \langle\textbf{n},\textbf{g}\rangle &= 
    &-\frac{d\tilde{H}_{1}}{dT} + \frac{\partial \tilde{H}_{0}}{\partial
      B}d^{B} + \frac{\partial \tilde{H}_{0}}{\partial I}d^{I} 
    = -\frac{d\tilde{H}_{1}}{dT} - \gamma B\frac{d\theta}{dT} - \gamma
    (3I^{r}-2B)\frac{d\phi}{dT}\nonumber\\ &=
    &-\frac{d\tilde{H}_{1}}{dT} - 3\gamma I^r\frac{d\phi}{dT} +
    2\gamma B\frac{d\chi_1}{dT} - 3\gamma B\frac{d\theta}{dT}, 
\end{eqnarray}
where we recall that $\chi_1=\theta+\phi$.

We can now explicitly integrate each of the terms in the
integrand~(\ref{integrand2}). From Eq.~(\ref{H1Btheta}), owing to the
fact that on the homoclinic orbits $B(\pm\infty)=0$, the first of
these yields
\begin{eqnarray}\label{int-term1}
    \int^{\infty}_{-\infty}\frac{d\tilde{H}_{1}}{dT}dT &=
    &-\frac{3\delta I^{r}h}{2}\left[\cos2\phi(\infty) -
    \cos2\phi(-\infty)\right]\nonumber\\  
    &= &-\frac{3\delta I^{r}h}{2}\left[\cos2(\phi_{0} +
    \Delta\phi/2) - \cos2(\phi_{0} - \Delta\phi/2)\right]\nonumber\\ 
    &= &3\delta I^{r}h\sin2\phi_{0}\sin\Delta\phi,
\end{eqnarray}
where we recall that $\Delta\phi = \phi(\infty) - \phi(-\infty)$. The
second term in~(\ref{integrand2}) immediately yields $-3\gamma I^r\Delta\phi$.
For the third term in~(\ref{integrand2}) we use Eq.~(\ref{psi dot}),
which on resonance yields
\begin{eqnarray}\label{int-term3}
    \int^{\infty}_{-\infty}B\frac{d\chi_1}{dT} dT &= &\frac{1 -
    \delta^{2}}{2\delta}\int^{\infty}_{-\infty}B^{2}dT\nonumber\\ 
    & = &(1 - \delta^{2})2\delta a^3
    \left(\frac{2p}{(p^{2} - q^{2})^{3/2}} \textmd{
    arctanh}\sqrt\frac{p-q}{p+q} + \frac{1}{q^{2} - p^{2}}\right)
    \equiv\Delta\sigma. 
\end{eqnarray}

For the fourth and last term in (\ref{integrand2}) we use Eq.~(\ref{B
  as theta}) and get 
\begin{equation}\label{int-term4}
    \int Bd\theta = I^{r}\theta + \frac{I^{r}(\delta^2-1)
    -2\delta\Omega_{1}} {(\delta-1)\sqrt{\delta^{2} -6\delta +1}}
    \arctan\left(\frac{\delta-1}{\sqrt{\delta^{2} -6\delta
    +1}}\tan\theta\right),  
\end{equation}
and after substituting the limits, using Eq.~(\ref{theta}), we get
\begin{eqnarray}\label{int-term4-limits}\nonumber
    \int Bd\theta &= &I^{r}\Delta\theta -2 \frac{I^{r}(\delta^2-1)
    -2\delta\Omega_{1}} {(\delta-1)\sqrt{\delta^{2} -6\delta +1}}
    \arctan\left(\frac{\delta-1}{\sqrt{\delta^{2} -6\delta
    +1}}
    \sqrt\frac{I^r(\delta-3)-\Omega_1}{I^r(1-\delta)+\Omega_1}\right)\\
  &\equiv &I^{r}\Delta\theta + \Delta\mu.  
\end{eqnarray}
After collecting all four terms we finally obtain
\begin{equation}\label{Melnikov}
    M(I^{r},\phi_{0}) = 
    -3\delta I^{r}h\sin2\phi_{0}\sin\Delta\phi -
    \gamma(3I^{r}\Delta\chi_{1} + 3\Delta\mu - 2\Delta\sigma).
\end{equation}
Except for the special case in which the phase difference $\Delta\phi$
is a multiple of $\pi$, the function $M(I^{r},\phi_{0})$ has simple
zeros as long as the relation
\begin{equation}\label{inter cond}
    \left|\frac{\gamma(3I^{r}\Delta\chi_{1} + 3\Delta\mu -
    2\Delta\sigma)}{3\delta I^{r}h\sin\Delta\phi}\right| < 1 
\end{equation}
is satisfied. If the system parameters satisfy this condition, every
simple zero of the Melnikov function corresponds to two symmetric (due
to the invariance $x,y\rightarrow -x,-y$) two-dimensional intersection
surfaces. The $\xi\rightarrow0$ limit of these surfaces contain orbits
whose explicit form is given by Eqs.~(\ref{orbit}), with their $I$ and
$\phi_{0}$ values satisfying the relation $M(I,\phi_{0}) = 0$, for $I$
close to $I^{r}$~\cite{kovacic2,kovacic3}. Thus, an unperturbed
heteroclinic orbit given by Eqs.~(\ref{orbit}), with $I = I^r$ and a
phase $\phi_0$ at time zero, can be made to persist under the
perturbation by setting the drive amplitude to the value
\begin{equation}
    h = \frac{\gamma(2\Delta\sigma - 3I^{r}\Delta\chi_{1} -
    3\Delta\mu)}{3\delta I^{r}\sin2\phi_{0}\sin\Delta\phi}. 
\end{equation}
We give numerical evidence of this in Sec.~\ref{silnikov}. Such orbits
surviving in the intersection of $W^{u}(\mathscr{M}_{\xi})$ and
$W^{s}(\mathscr{M}_{\xi})$ may leave the stable manifold
$W^{s}(\mathscr{M}_{\xi})$ in forward time, and the unstable manifold
$W^{u}(\mathscr{M}_{\xi})$ in backward time, through the low boundary
at $I=\Omega_1/(\delta-1)$, since these manifolds are only locally
invariant~\cite{kovacic3}. However, the analysis we perform below
allows us to find surviving homoclinic orbits that are contained in
the intersection of $W^{u}(\mathscr{M}_{\xi})$ and
$W^{s}(\mathscr{M}_{\xi})$.

\section{DYNAMICS NEAR RESONANCE}
\label{resonance}

After having calculated the Melnikov integral at $I = I^{r}$, we
proceed to examine the dynamics on $\mathscr{M_{\xi}}$ near this
resonance.  The equations that describe the dynamics on
$\mathscr{M_{\xi}}$ are obtained by setting $B = 0$ in
Eqs.~(\ref{per3}) and (\ref{per4}),
\begin{subequations}
\label{M_xi}
\begin{eqnarray}
 \frac{dI}{dT} & = &   - \xi3I(h\delta \sin2\phi + \gamma),\\
\frac{d\phi}{dT} & = & - \frac{\Omega}{4} + \delta I -
\xi\frac{3h\delta}{2}\cos2\phi. 
\end{eqnarray}
\end{subequations}
To investigate the slow dynamics, which is induced by the perturbation
on $\mathscr{M_{\xi}}$ near resonance, we follow Kova\v{c}i\v{c} and
Wiggins~\cite{kovacic1,kovacic3} and introduce a slow variable $I =
I^{r} + \sqrt{\xi}\rho$ into Eq.~(\ref{M_xi}), along with a slow time
scale $\tau = \sqrt{\xi}T$, and obtain
\begin{subequations}
\label{scaled}
\begin{eqnarray}
 \frac{d\rho}{d\tau} & = &  - 3(I^{r} +
 \sqrt{\xi}\rho)(h\delta\sin2\phi + \gamma) ,\\ 
\frac{d\phi}{d\tau} & = &\delta\rho - \sqrt{\xi}\frac{3h\delta}{2}\cos2\phi.
\end{eqnarray}
\end{subequations}
The leading terms in Eqs.~(\ref{scaled}), independent of $\xi$, yield
\begin{subequations}
\label{outer}
\begin{eqnarray}
 \frac{d\rho}{d\tau} & = & - 3I^{r}(h\delta\sin(2\phi) + \gamma) =
 -\frac{\partial\mathscr{H}(\rho,\phi)}{\partial\phi},\\ 
\frac{d\phi}{d\tau} & = &\delta\rho =
 \frac{\partial\mathscr{H}(\rho,\phi)}{\partial\rho}, 
\end{eqnarray}
\end{subequations}
where
\begin{equation}\label{slowH}
    \mathscr{H}(\rho,\phi) =   \frac{1}{2}\delta\rho^{2} -
    \frac{3}{2}h\delta I^{r}\cos(2\phi) + 3\gamma I^{r}\phi 
\end{equation}
is a rescaled Hamiltonian that governs the slow dynamics on
$\mathscr{M_{\xi}}$ close to resonance.

\begin{figure}
\begin{center}
\centering
  \subfigure[\ $\xi=0$]{
  \includegraphics[width = 0.46\textwidth]{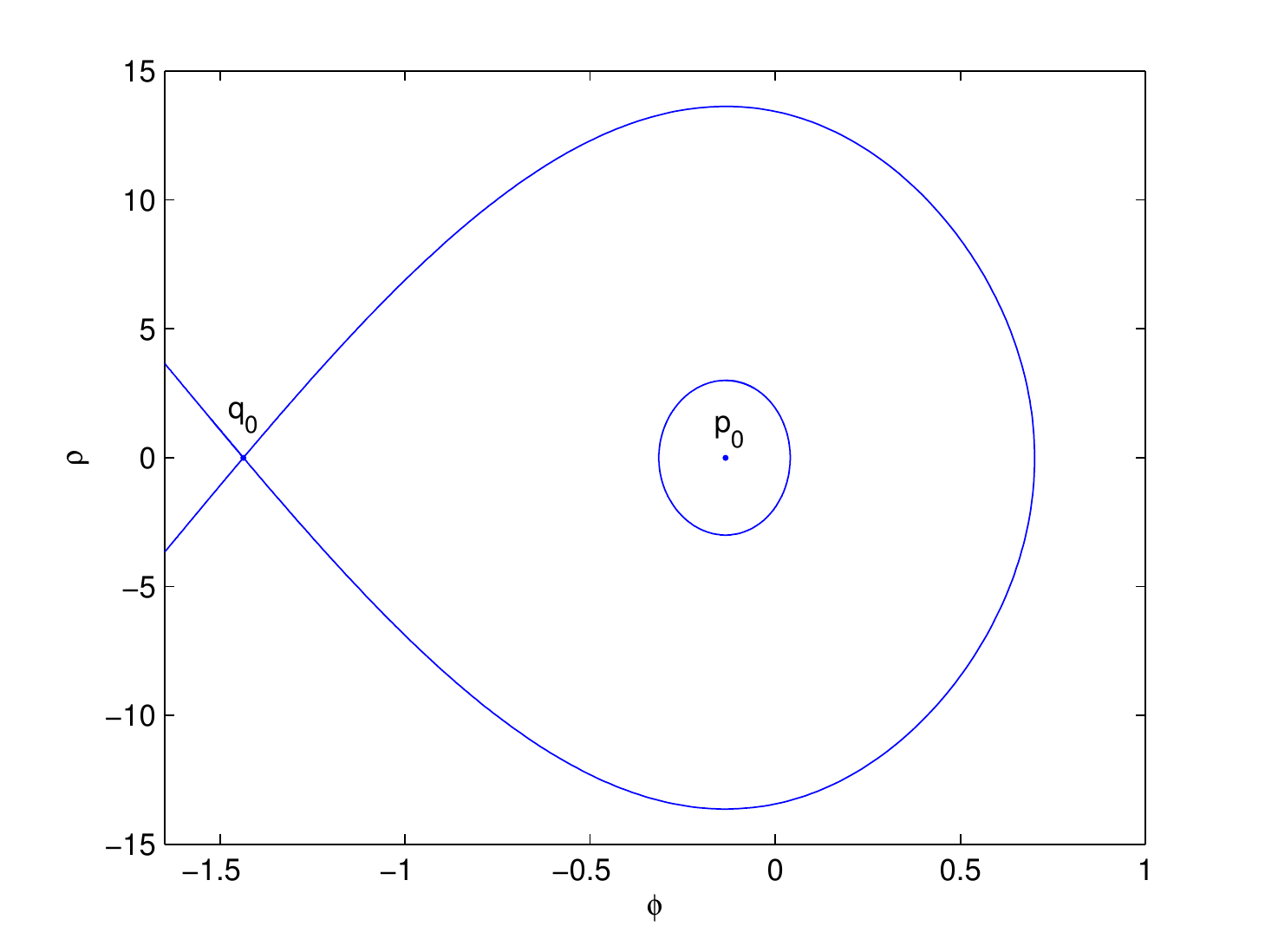}}
    \subfigure[\ $\xi=1$]{
   \includegraphics[width = 0.46\textwidth]{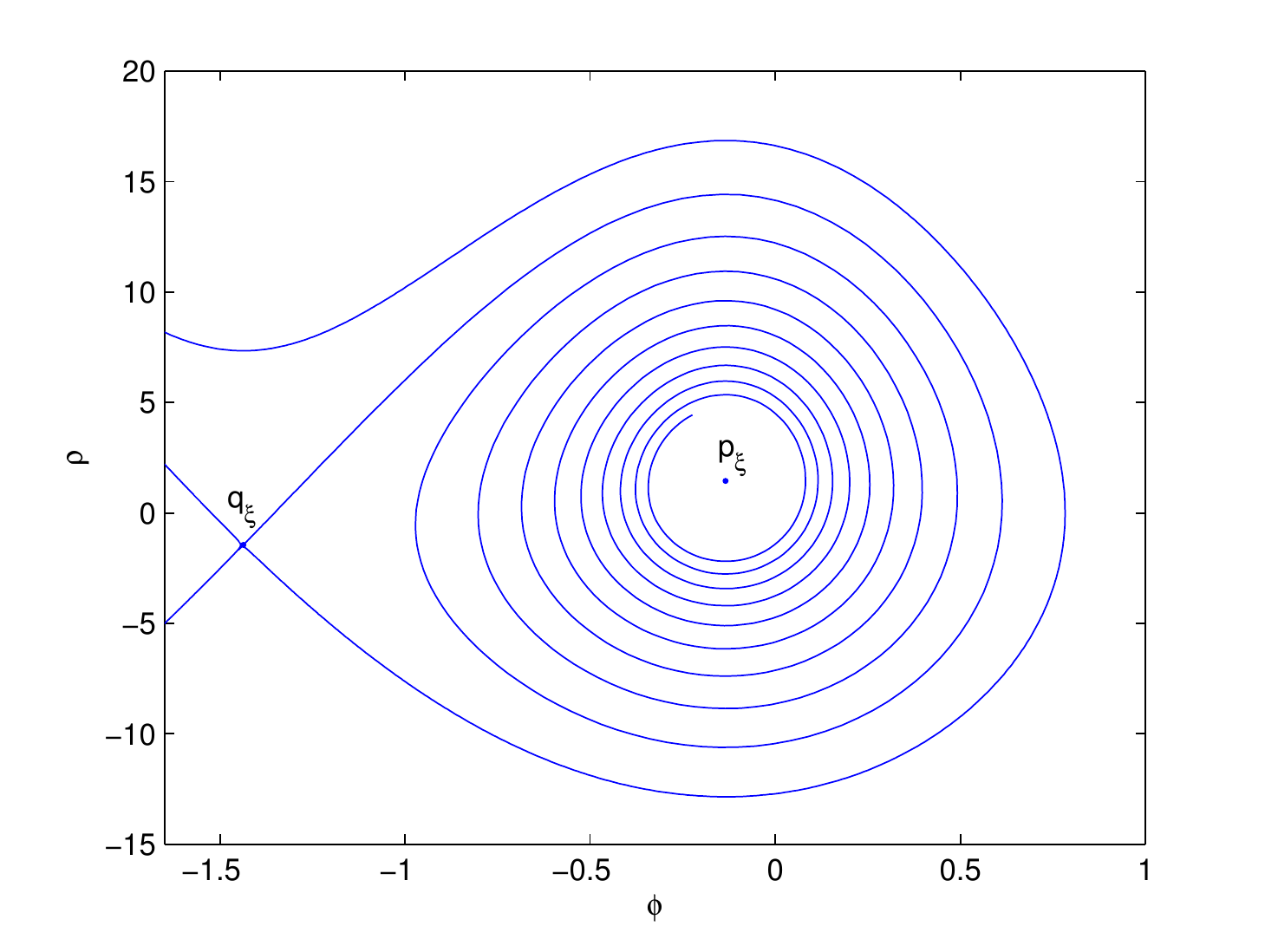}}
   \end{center}
   \caption{\label{phase} (a) Numerical phase portraits of
     Eqs.~(\ref{scaled}) with $\xi = 0$ [or equivalently,
     Eqs.~(\ref{outer})], showing a saddle and a center. (b) Numerical
     phase portraits of Eqs.~(\ref{scaled}) with $\xi = 1$, showing
     that the saddle remains a saddle but that the center becomes a
     sink, with their $\rho$ coordinates shifted slightly down and up,
     respectively.  The parameters are $\delta = 2,\Omega = 400,h =
     1,b = 0.2649,\gamma = hb\delta$. }
\end{figure}

Fig.~\ref{phase}(a) shows the phase portrait of Eqs.~(\ref{outer}),
which contains a saddle $q_{0}$ at $(\rho = 0,\phi = \phi_{s} =
[\arcsin b - \pi]/2)$, and a center $p_{0}$ at $(\rho = 0,\phi =
\phi_{c} = -[\arcsin b]/2)$, where $b\equiv\gamma/h\delta$. The
fixed points of Eqs.~(\ref{scaled}) that contain the additional
$O(\sqrt\xi)$ terms are $q_{\xi} = (-\rho_{\xi},\phi_{s})$ and
$p_{\xi} = (\rho_{\xi},\phi_{c})$, where $\rho_{\xi} =
\sqrt{\xi}3h\sqrt{1 - b^{2}}/2$.  For small positive $\xi$, a linear
analysis of these fixed points reveals that $q_{\xi}$ is still a
saddle but that $p_{\xi}$ is a sink, as shown in Fig.~\ref{phase}(b).
The fixed points of the full equations~(\ref{M_xi}) near $I=I^{r}$ are
the same saddle and sink, located at $(I = I^-,\phi = \phi_{s})$ and
$(I = I^+,\phi = \phi_{c})$, respectively, where $I^{\pm} =
I^{r}\pm\sqrt{\xi}\rho_{\xi}$.

The scaled equations (\ref{outer}) provide an estimate for the basin
of attraction of the sink, which is the area confined within the
homoclinic orbit connecting the saddle $q_{0}$ to itself, shown in
Fig.~\ref{phase}(a). Recall that the dynamics on the unperturbed
annulus $\mathscr{M}$ is composed of simple one-dimensional flows,
which on resonance turn into a circle of fixed points. Upon adding the
small perturbation, two of these fixed points persist in an interval
of length $\pi$, and the phase space contains two-dimensional flows.
Of particular interest is the basin of attraction of the sink, because
a homoclinic orbit to a fixed point of this type offers a mechanism
for producing chaotic motion. This mechanism, which results from the
existence of a homoclinic trajectory to a saddle-focus fixed point,
was described by \v{S}ilnikov~\cite{silnikov}.  Obtaining an estimate
for the basin of attraction of the sink, allows us to pick-out the
trajectories satisfying \v{S}ilnikov's theorem, which we do in the
following section.

\section{A HOMOCLINIC CONNECTION TO THE SINK $p_{\xi}$}
\label{silnikov}

We are finally in a position to show the existence of an orbit
homoclinic to the sink $p_{\xi}$. Note that for a particular set of
parameters the existence of such an orbit implies the existence of
another symmetric orbit due to the invariance $(x,y)\rightarrow
(-x,-y)$. To achieve this, we first show that there exists a
homoclinic orbit that approaches $p_{\xi}$ asymptotically backward in
time, and approaches the perturbed annulus $\mathscr{M}_{\xi}$
asymptotically forward in time. We then estimate the conditions under
which the perturbed counterpart of the point, which is reached on
$\mathscr{M}$ forward in time in the unperturbed system, lies within
the basin of attraction of the sink $p_{\xi}$ on $\mathscr{M}_{\xi}$.
This gives us an estimate for the possibility of obtaining a
\v{S}ilnikov orbit that connects the sink back to itself.

The first step is done by finding the conditions for which the
Melnikov function $M(I^{r},\phi_{0} = \phi_{c} + \Delta\phi/2)$ has
simple zeros. We substitute $\phi_{0} = \phi_{c} + \Delta\phi/2$ into
the first term in Eq.~(\ref{Melnikov}), and recall that
$\sin2\phi_c=-b$, to get
\begin{equation}\label{subs phi}
  \sin2\phi_{0}\sin\Delta\phi = \frac{1}{2}\left[\sqrt{1 - b^{2}} (1 -
  \cos2\Delta\phi) - b\sin2\Delta\phi\right]. 
\end{equation}
By substituting (\ref{subs phi}) into the Melnikov function
(\ref{Melnikov}) and equating it to zero we obtain the equation
\begin{eqnarray}
  3I^{r}\left[\sqrt{1 - b^{2}} (1 -
  \cos2\Delta\phi) - b\sin2\Delta\phi\right] 
  + 2b(3I^{r}\Delta\chi_{1} + 3\Delta\mu - 2\Delta\sigma) = 0,
\end{eqnarray}
from which we extract an explicit expression for the condition on $b$,
ensuring the existence of an orbit that asymptotes to $p_{\xi}$
backwards in time, and to $\mathscr{M}_{\xi}$ forward in time,
\begin{equation}\label{b}
    |b| = \frac{1 -
    \cos2\Delta\phi}{\sqrt{\left(\frac{4}{3I^{r}}\Delta\sigma +
    \sin2\Delta\phi - 2\Delta\chi_{1} -
    \frac{2\Delta\mu}{I^{r}}\right)^{2} + (1 -
    \cos2\Delta\phi)^{2}}}.
\end{equation}

\begin{figure}
  \includegraphics[width = 0.7\textwidth]{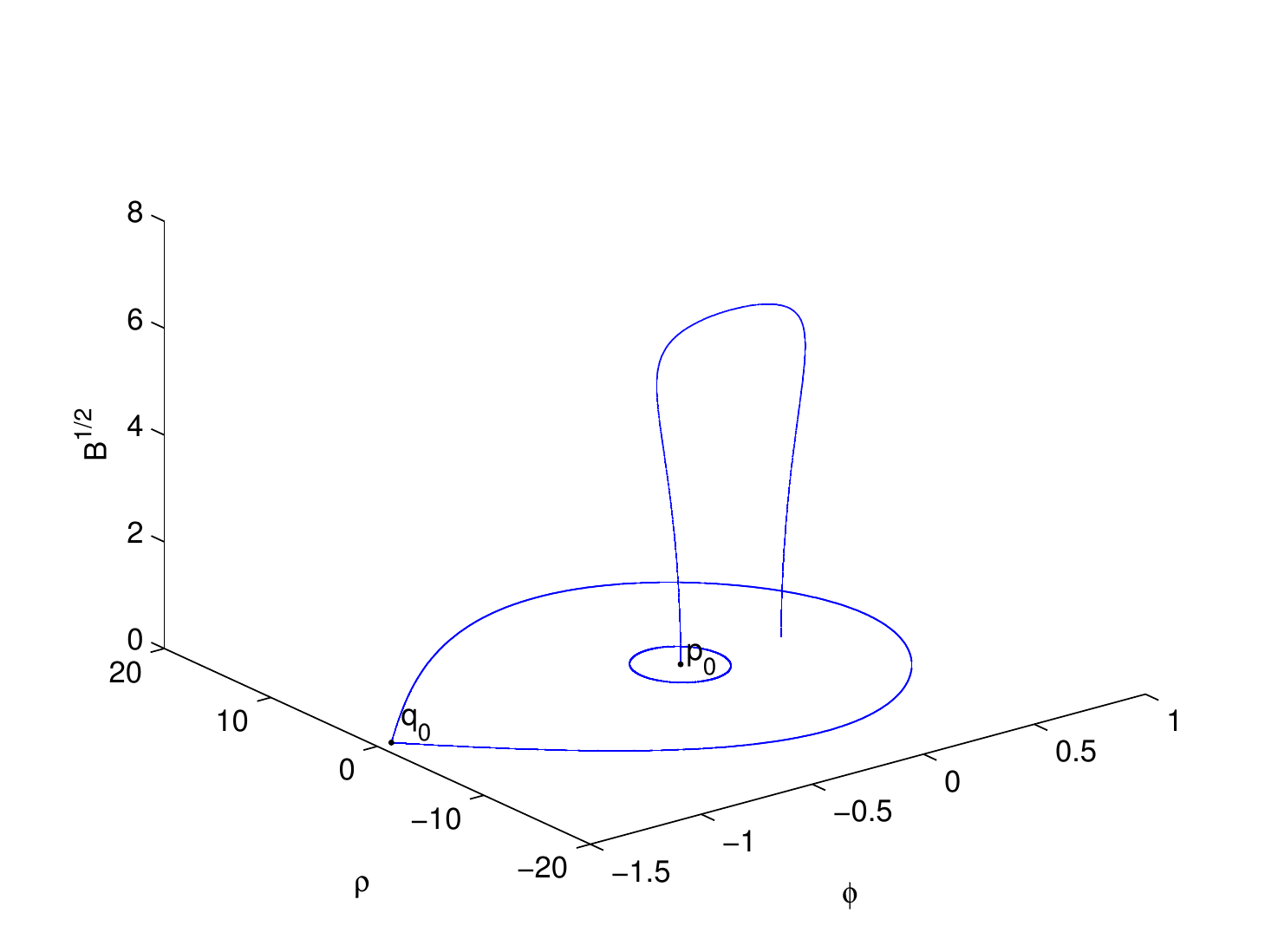}
  \caption{\label{unp_super} The heteroclinic orbit given by
    Eqs.~(\ref{orbit}) with $I = I^{r}$, superimposed with the phase
    portrait of the unperturbed scaled system on $\mathscr{M}_{\xi}$
    near resonance, given by Eqs.~(\ref{outer}). The parameters are
    the same as in Fig.~\ref{phase}(a), with $\Omega_{1} = 21.32$.
    For these parameters $b = 0.2949$ according to Eq.~(\ref{b}), so
    we fix $h = 1$ and $\gamma = \delta b h$ in Eqs.~(\ref{outer}).
    This value of $b$ sets $\phi(-\infty) = \phi_{c} = - 0.1341$, and
    as can be seen from the figure $\phi_{s} < \phi(\infty) = \phi_{c}
    + \Delta\phi < \phi_m$.}
\end{figure}

Next, we wish to find an approximate condition, ensuring that this
orbit approaches $p_{\xi}$ as $T\rightarrow\infty$. To do so we find
the condition for which the \emph{unperturbed} heteroclinic orbit, which
asymptotes to $p_{0}$ as $T\rightarrow -\infty$, returns back to a
point on the circle of fixed points that is inside the homoclinic
separatrix loop connecting the saddle $q_{0}$ to
itself~\cite{kovacic1}. Such an orbit is shown in
Fig.~\ref{unp_super}. This condition is formulated in terms of the
difference $\Delta\phi$ between the asymptotic values of the angular
variable $\phi$ as
\begin{equation}\label{condition}
    \phi_{s} < \phi_{c} + \Delta\phi < \phi_m,
\end{equation}
where $\phi_m$ is the maximal value of $\phi$ on the homoclinic orbit,
connecting the saddle $q_{0}$ to itself. Since the Hamiltonian is
conserved along an orbit, $\phi_m$ satisfies the equation
\begin{eqnarray}\label{phi-m}
    0& = &\mathscr{H}(0,\phi_m) - \mathscr{H}(0,\phi_{s})\nonumber\\
    &= &3I^{r}h\delta\left[\frac{1}{2}\sqrt{1-b^2}
    + \frac{1}{2}\cos2\phi_m - b\left(\phi_m + \frac{\pi}{2}
    - \frac{1}{2}\arcsin b\right)\right], 
\end{eqnarray}
whose roots are found numerically to obtain $\phi_m$.

\begin{figure}
\begin{center}
\centering
\subfigure[]{
  \includegraphics[width = 0.48\textwidth]{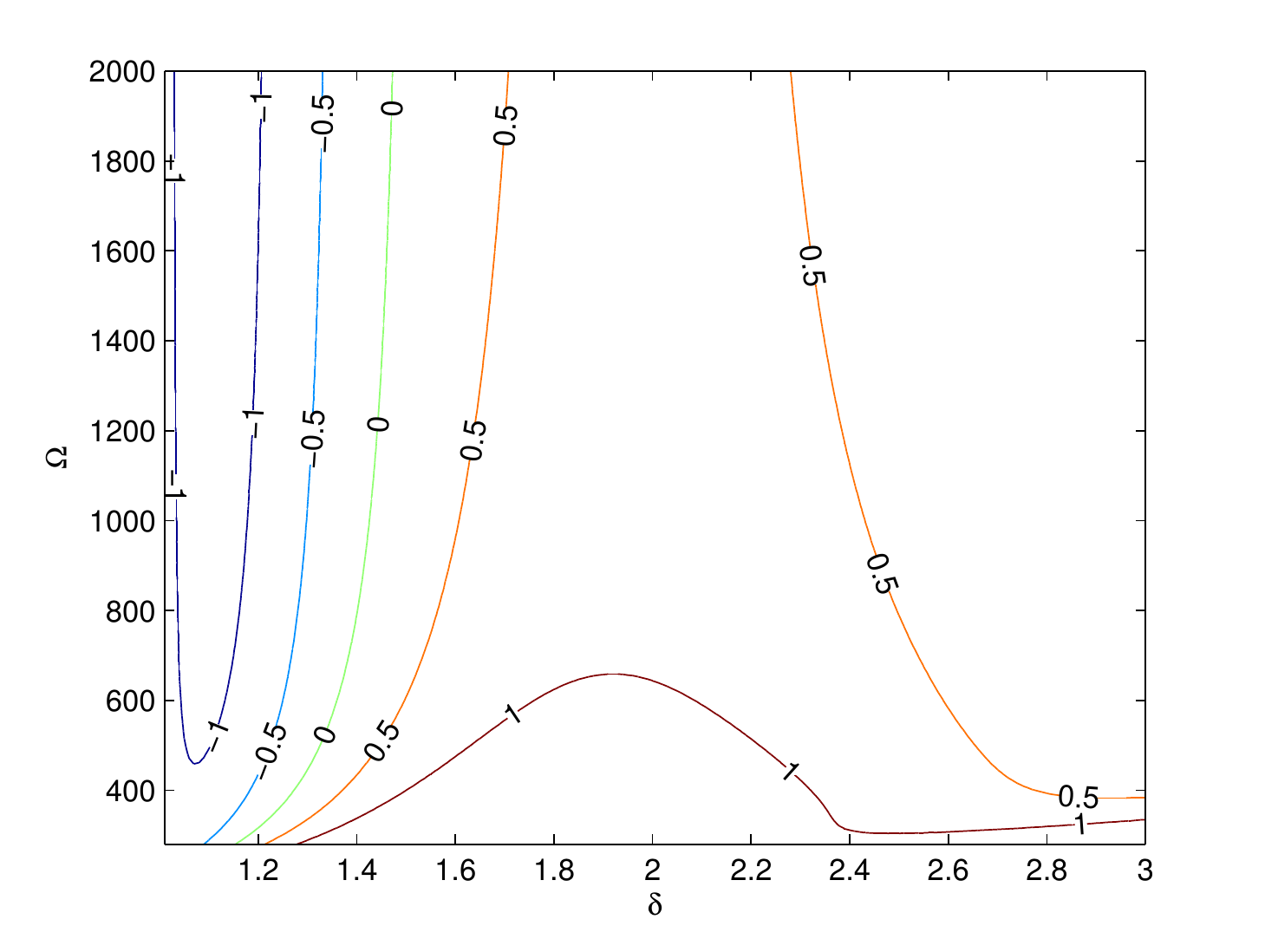}
}
\subfigure[]{
  \includegraphics[width = 0.47\textwidth]{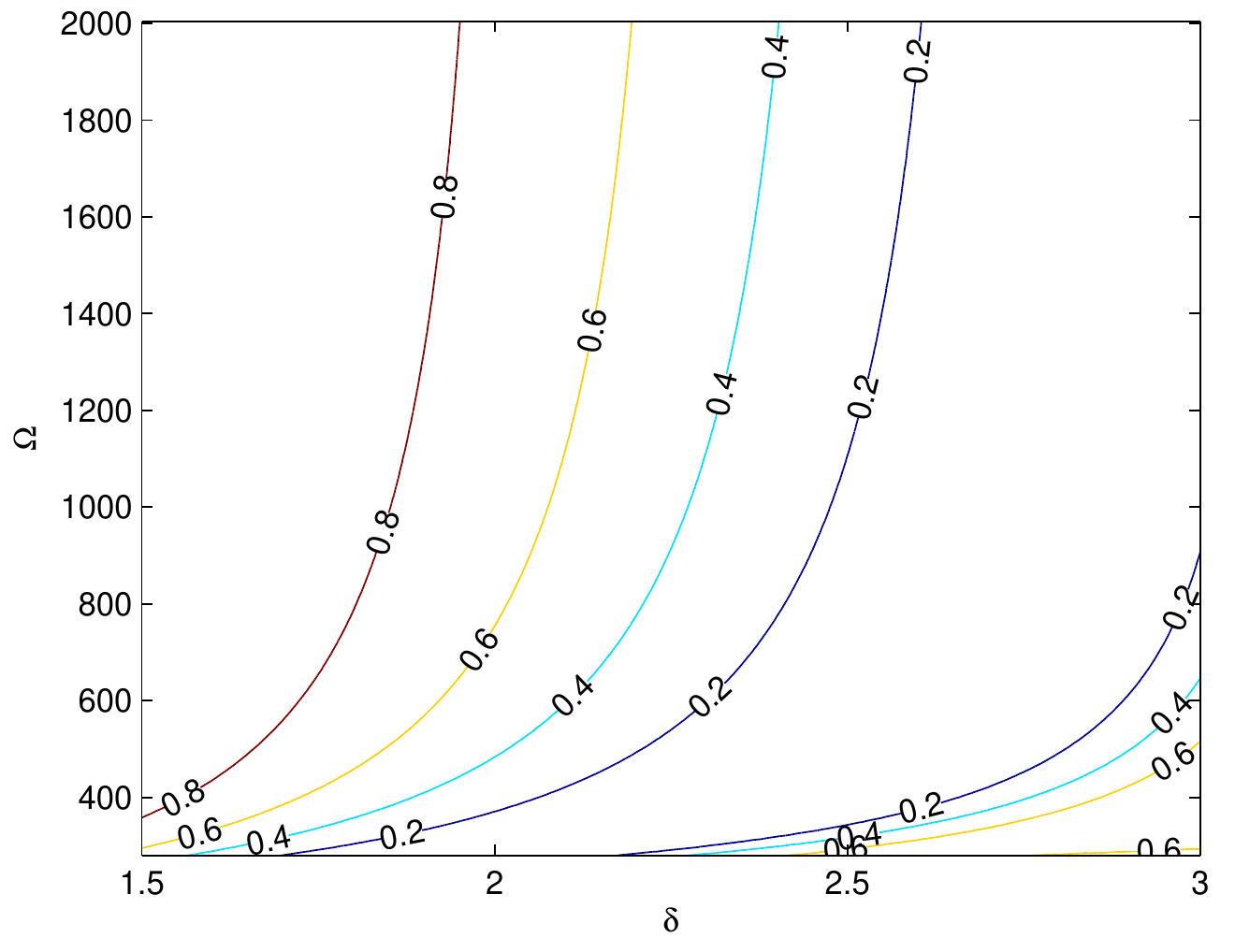}
}
\end{center}
  \caption{\label{b-condition} (Color online) (a) Contour plot
    of the left-hand side of the inequality~(\ref{positive b}).  In
    the displayed range of $\Omega$, for $\delta\gtrsim 1.5$, this
    function is positive and the coefficient $\gamma$ represents
    energy dissipation. For fixed $\epsilon = 0.01$ and $1 < \delta <
    3$, the scaled frequency $\Omega_{1}$ reaches values of $0 <
    \Omega_{1} < 28$. (b) The ratio $b=\gamma/h\delta$, given by
    Eq.~(\ref{b}), as a function of $\Omega$ and $\delta$ $(\epsilon =
    0.01)$. Here $1.5 < \delta < 3$ and $b$ is positive.}
\end{figure}

\begin{figure}
\begin{center}
\centering
\subfigure[]{
  \includegraphics[width = 0.48\textwidth]{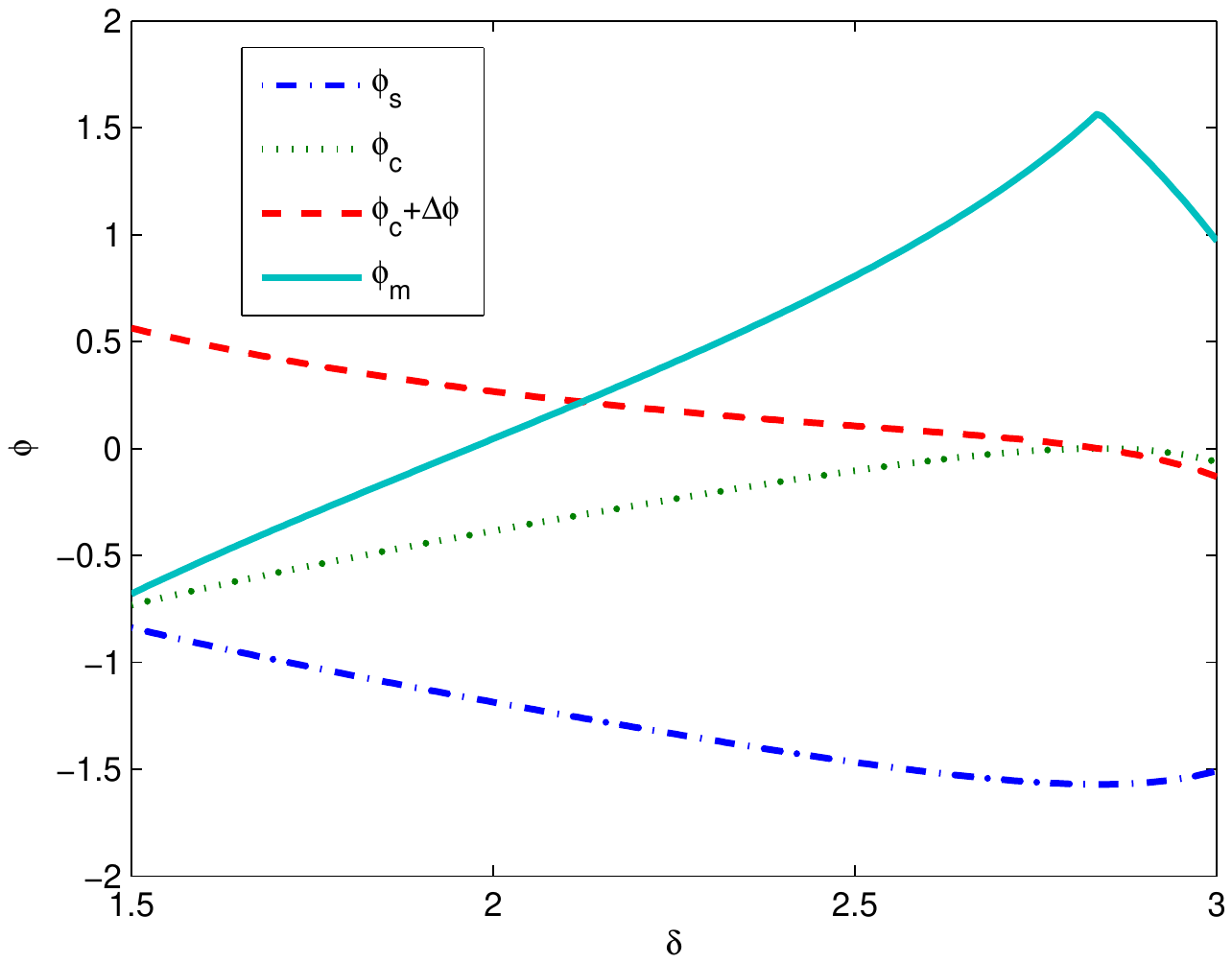}}
\subfigure[]{
  \includegraphics[width = 0.48\textwidth]{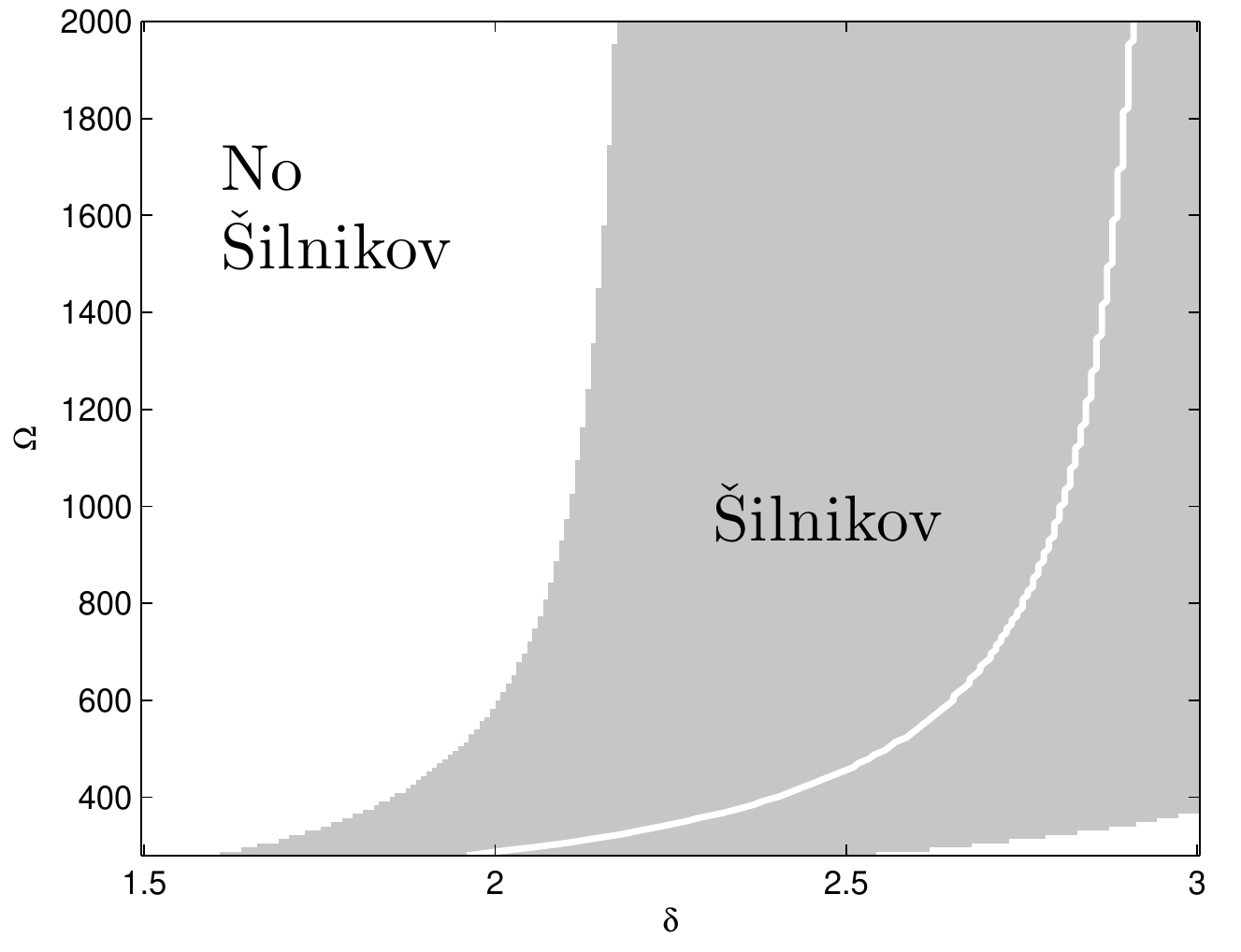}}
\end{center}
\caption{\label{final conditions} (a) (Color online) The values of
  $\phi_{s}$, $\phi_{c}$, $\phi_{c} + \Delta\phi$, and $\phi_m$ as
  functions of $\delta$, for $\Omega = 1135.64$. For $\delta >
  2.12$ the condition~(\ref{condition}) is satisfied and orbits
  homoclinic to the sink $p_{\xi}$ exists, except when $\Delta\phi
  = 0$. (b) Parameter values for which the
  condition~(\ref{condition}) is satisfied are indicated in gray.
  The white line inside the gray area corresponds to $\Delta\phi =
  0$, where the theory does not apply. In both figures $\epsilon =
  0.01$.}
\end{figure}

%\section{DISCUSSION -- PHYSICAL PARAMETERS \& NUMERICAL VERIFICATION}

\begin{figure}[!htb]
\begin{center}
\centering
\subfigure[]{
  \includegraphics[width = 0.46\textwidth]{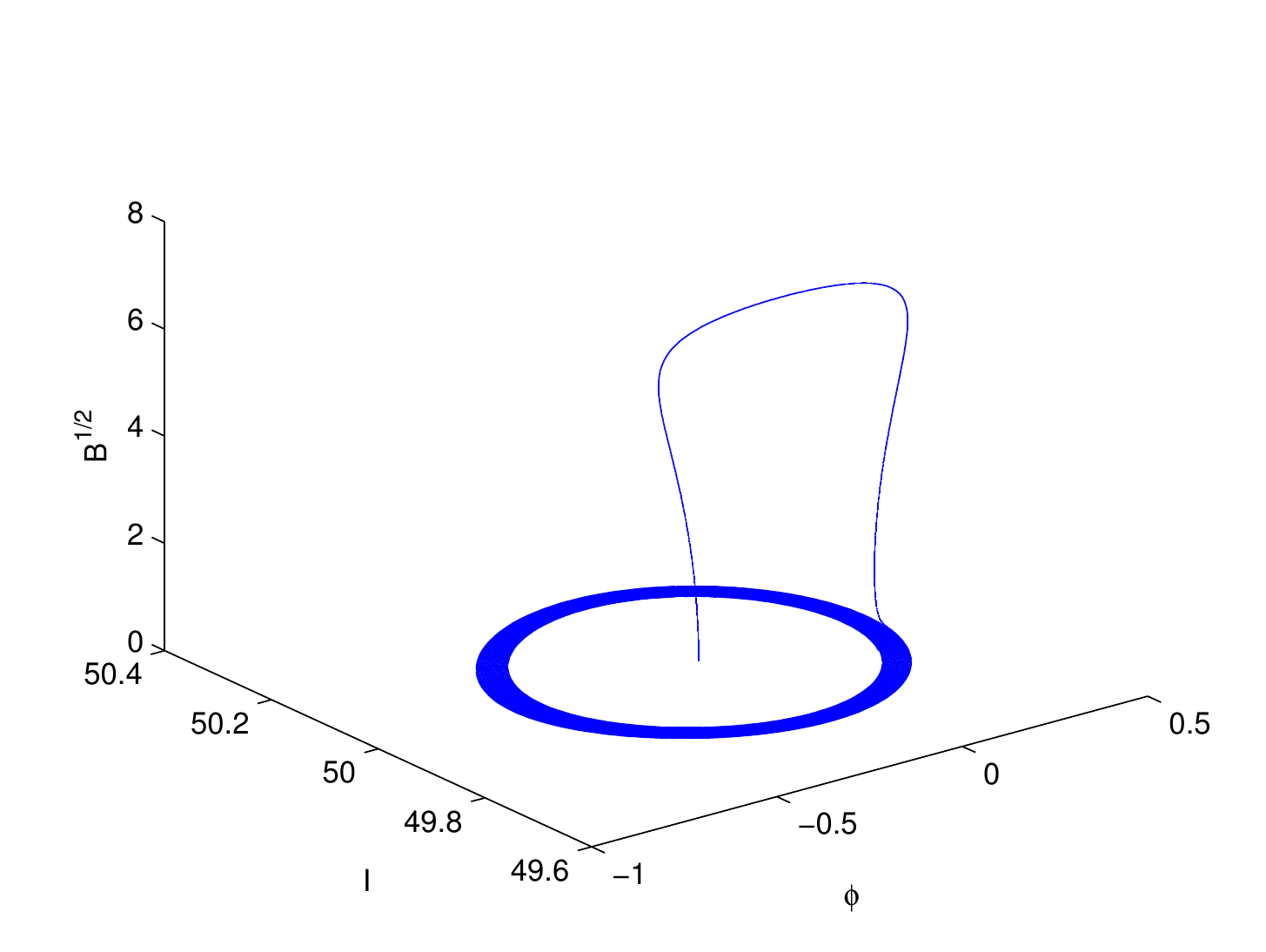}}
\subfigure[]{
  \includegraphics[width = 0.46\textwidth]{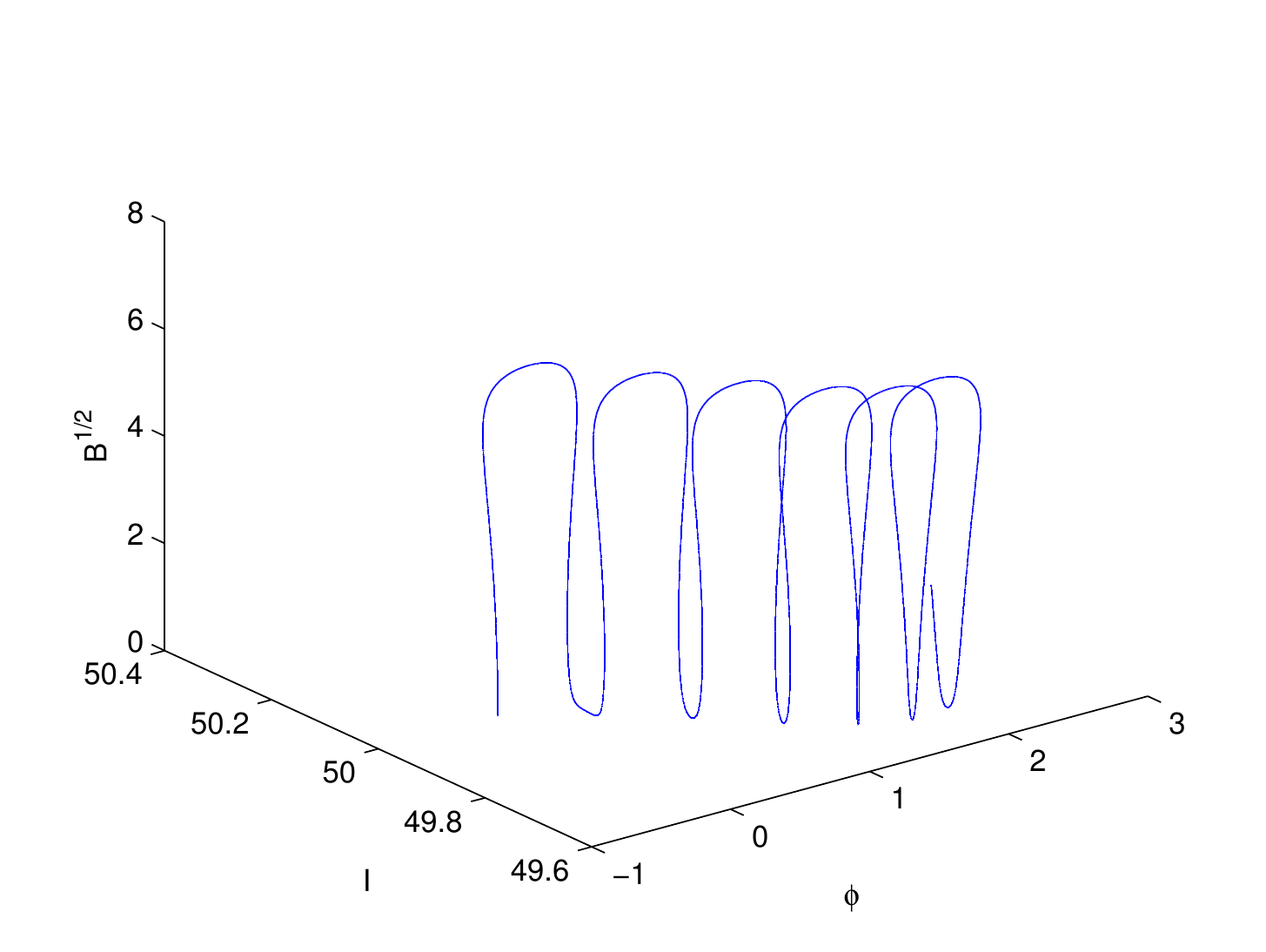}}\\
\subfigure[]{
  \includegraphics[width = 0.46\textwidth]{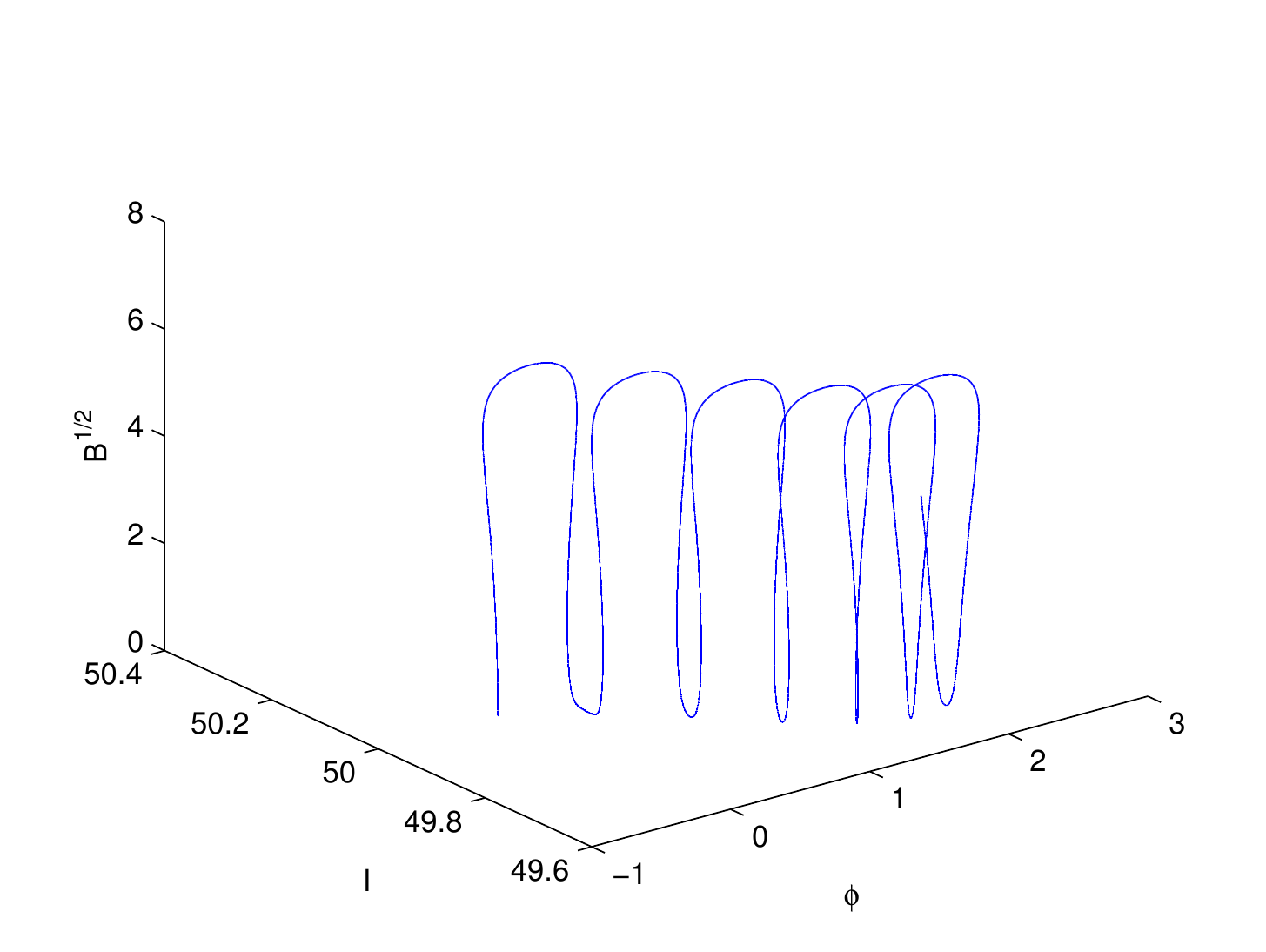}}
\subfigure[]{
  \includegraphics[width = 0.46\textwidth]{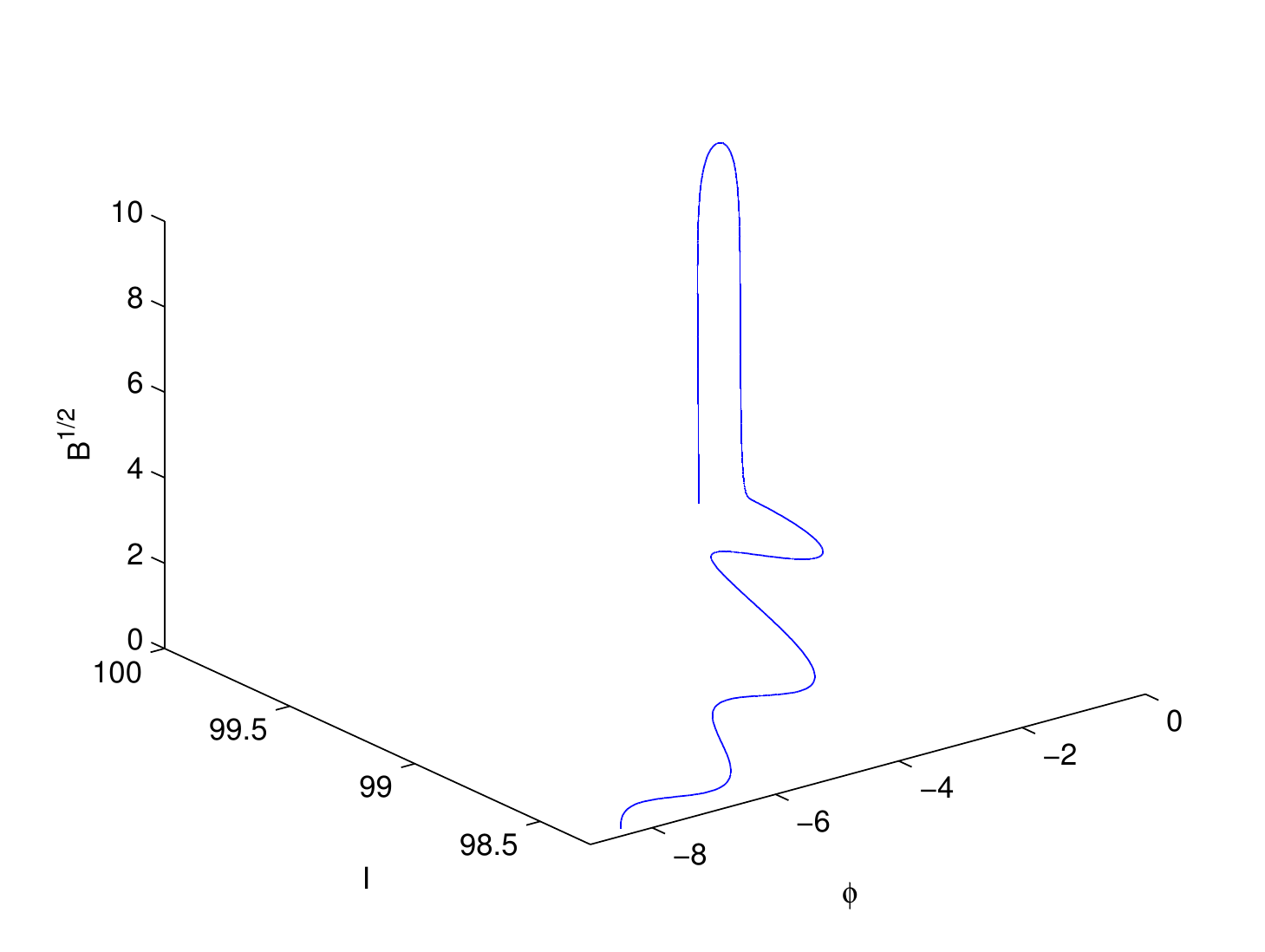}}
\end{center}
\caption{\label{numerics}Results of our numerical scheme for $\xi =
  0.001$ and the rest of the parameters as in Fig.~\ref{unp_super}.
  (a) An illustration of the \v{S}ilnikov orbit that is obtained for
  $\gamma = hb\delta$ (The $\xi\rightarrow0$ limit of this orbit is
  shown in Fig.~\ref{unp_super}).  For (b) $\gamma = h(b -
  0.0003)\delta$, and (c) $\gamma = h(b + 0.0002)\delta$, and we see
  that the orbit does not get close enough to $\mathscr{M}_{\xi}$ in
  order to meet our cutoff criterion for being homoclinic to it.  (d)
  For $\Omega = 800$, we obtain an orbit that approaches
  $\mathscr{M}_{\xi}$ by setting the appropriate value of $\gamma$ $(b
  = 0.6178)$, however, this orbit does not asymptote to the
  saddle-focus, in agreement with Fig.~\ref{final conditions} (b). In
  this simulation, the orbit leaves $\mathscr{M}_{\xi}$ through its
  boundary at $I = \Omega_1/(\delta-1)$ and eventually $I\rightarrow0$
  and the motion dies out. }
\end{figure}

Eqs.~(\ref{b}) and (\ref{condition}) define conditions for the
existence of orbits homoclinic to the sink $p_{\xi}$. We wish to
relate these results to the actual physical parameters of the coupled
resonators. Recall that $\delta$ sets the value of the electrostatic
coupling coefficient $D = 2(\delta^{2} - 1)/(3\delta^{2} - 1)$. The
scaled frequency $\Omega_{1}$ is then given by $\Omega_{1} =
(\omega_{1} - \omega_{2})/2\epsilon = (\sqrt{1 - D/2} - \sqrt{1 -
  3D/2})/2\epsilon$, so by fixing $\epsilon$ it is also determined by
$\delta$. The ratio $b=\gamma/h\delta$, between the damping
coefficient and the drive amplitude, has to be positive in order for
the damping coefficient $\gamma$ to be positive and have the standard
physical meaning of energy dissipation. The ratio $b$ is positive if
the inequality
\begin{equation}\label{positive b}
    \frac{4}{3I^{r}}\Delta\sigma + \sin2\Delta\phi - 2\Delta\chi_{1}
    - \frac{2\Delta\mu}{I^{r}} > 0 
\end{equation}
is satisfied. We plot the left-hand side of this inequality as a
function of $\Omega$ and $\delta$ in Fig.~\ref{b-condition}(a), and
find that it is positive if $\delta\gtrsim 1.5$. Consequently we plot
the ratio $b$ in Fig.~\ref{b-condition}(b) for $1.5 < \delta < 3$.
This value of $b$ then determines the $\phi$ values of the fixed
points of Eq.~(\ref{outer}), which are shown in Fig.~\ref{final
  conditions}(a), along with $\phi_{c} + \Delta\phi$ and $\phi_m$ for
a particular value of $\Omega$. The parameter values for which these
$\phi$ values satisfy the condition~(\ref{condition}) are displayed in
Fig.~\ref{final conditions}(b), which outlines the values of the
electrostatic coupling and parametric driving frequency, for which
orbits homoclinic to the sink $p_{\xi}$ exist. We note that
\v{S}ilnikov orbits were also found in other two-mode parametrically
driven systems~\cite{haller95,feng931,fandw,zhang}, however, slightly
different equations were studied, resulting in different phase space
dynamics for the unperturbed system as well as different
perturbations.

Finally, we wish to verify our calculations by a numerical solution of
the ODEs~(\ref{per}). The difficulty in producing a \v{S}ilnikov orbit
in these equations is that the linearized growth rates of the
saddle-focus fixed point---a saddle on the $(B,\theta)$ plane and a
focus on the perturbed annulus $\mathscr{M}_{\xi}$---are $O(\xi)$ in
directions tangent to $\mathscr{M}_{\xi}$, so the orbit has to spend a
lot of time near $\mathscr{M}_{\xi}$ in order to spiral around the
saddle-focus.  However, the linearized growth rates of this fixed
point in directions transverse to $\mathscr{M}_{\xi}$, are $O(1)$, so
a small and inevitable numerical error would deflect the orbit away
from $\mathscr{M}_{\xi}$. To avoid this problem we solve the
ODEs~(\ref{per}) using a cutoff criterion. We initiate the numerical
solution with $B\ll1$, and the exact coordinates of the sink on
$\mathscr{M}_{\xi}$, $(I = I^{ + },\phi = \phi_{c})$. The orbit
initially flows away from $\mathscr{M}_{\xi}$ and later turns around
and approaches it. If on its way back towards $\mathscr{M}_{\xi}$, the
orbit approaches it close enough to satisfy $B < \xi/1000$, we set
$dB/dT = d\theta/dT = 0$ in Eq.~(\ref{per}), thus restricting the
motion to be tangent to $\mathscr{M}_{\xi}$. This numerical scheme
allows us to verify our predictions, because as shown in
Fig.~\ref{numerics}, only when the damping coefficient is equal to
$\gamma = hb\delta$ ($\pm\sim0.1\%$), with $b$ given by Eq.~(\ref{b}),
is our cutoff criterion for eliminating the motion transverse to
$\mathscr{M}_{\xi}$ satisfied. Furthermore, as shown in
Figs.~\ref{numerics}(a) and (d), among the orbits that satisfy our
cutoff criterion, only the ones that satisfy the
condition~(\ref{condition}) asymptote to the saddle-focus.

Owing to a theorem of \v{S}ilnikov~\cite{kovacic1}, the existence of
orbits homoclinic to a saddle-focus fixed point in Eq.~(\ref{per})
implies that these equations contain chaotic motion in the sense of
horseshoes in their dynamics.

\section{SUMMARY}

We have studied the origin of chaotic dynamics, and provided
conditions for its existence, in a case of two parametrically-driven
nonlinear resonators. This was achieved by applying a method of
Kova\v{c}i\v{c} and Wiggins on transformed amplitude equations that
were derived from the equations of motion, which model an actual
experimental realization of coupled nanomechanical resonators. We
considered the amplitude of the drive and the damping to be small
perturbations and obtained explicit expressions for orbits homoclinic
to a two-dimensional invariant annulus in the unperturbed equations.
At resonance, we were able to calculate the Melnikov integral
analytically, and provide a primary condition for having homoclinic
orbits in the full, perturbed equations. By further studying the
effects of perturbations on the invariant annulus near resonance, we
found a secondary condition for the existence of orbits homoclinic to
a fixed point of a saddle-focus type. We used a numerical scheme to
verify our theoretical predictions. Such \v{S}ilnikov homoclinic
orbits give rise to a particular type of horseshoe chaos, which can be
expected in the dynamics of the full system for parameter values in
the vicinity of those presented here.

\section*{Acknowledgments}

EK and RL wish to thank Mike Cross and Steve Shaw for fruitful
discussions. This work was supported by the U.S.-Israel Binational
Science Foundation (BSF) through Grant No. 2004339, by the
German-Israeli Foundation (GIF) through Grant No. 981-185.14/2007, and
by the Israeli Ministry of Science and Technology.

\bibliography{chaos}

\end{document}